\documentclass[10pt, journal,compsoc]{IEEEtran}

\ifCLASSOPTIONcompsoc
  \usepackage[nocompress]{cite}
\else
  \usepackage{cite}
\fi

\usepackage{balance}
\usepackage{array,enumerate}
\usepackage{multirow}
\usepackage{graphicx}
\usepackage[linesnumbered,ruled,vlined]{algorithm2e}
\usepackage{float}
\SetArgSty{textup}
\usepackage{amssymb}
\usepackage{amsmath}
\usepackage{url}
\usepackage{threeparttable}
\usepackage{scrextend}
\usepackage{bm}
\usepackage{tikz}
\usepackage{algpseudocode}
\usepackage{ragged2e}
\usepackage{courier}
\usepackage{array}
\usepackage{enumitem} 

\newcolumntype{L}[1]{>{\raggedright\let\newline\\\arraybackslash\hspace{0pt}}m{#1}}
\newcolumntype{C}[1]{>{\centering\let\newline\\\arraybackslash\hspace{0pt}}m{#1}}
\newcolumntype{R}[1]{>{\raggedleft\let\newline\\\arraybackslash\hspace{0pt}}m{#1}}

\SetCommentSty{mycommfont}

\let\oldnl\nl
\newcommand{\nonl}{\renewcommand{\nl}{\let\nl\oldnl}}
\newcommand{\subcaption}[1]{\centerline{{\scriptsize
  #1}}\vspace{10pt}}
\newlength{\minipagewidth}
\newlength{\figurewidthFour} 
\makeatletter
\newcommand{\thickhline}{%
    \noalign {\ifnum 0=`}\fi \hrule height 1.2pt
    \futurelet \reserved@a \@xhline
}
\newcolumntype{"}{@{\hskip\tabcolsep\vrule width 1pt\hskip\tabcolsep}}
\makeatother
\begin{document}

\title{GraphMP: I/O-Efficient Big Graph Analytics on a Single Commodity Machine}
\author{
    Peng~Sun, Yonggang~Wen, Ta~Nguyen~Binh~Duong and Xiaokui~Xiao
    \IEEEcompsocitemizethanks{
        \IEEEcompsocthanksitem
        Peng Sun, Yonggang Wen and Ta Nguyen Binh Duong are with School of Computer Science and Engineering, Nanyang Technological University, Singapore. 
        Email: \{sunp0003, ygwen, donta\}@ntu.edu.sg
        \IEEEcompsocthanksitem
        Xiaokui Xiao is with School of Computing, National  University of Singapore, Singapore. Email: xkxiao@nus.edu.sg
    }
}

\markboth{}%
{Shell \MakeLowercase{\textit{et al.}}: Bare Demo of IEEEtran.cls for Journals}

\IEEEtitleabstractindextext{

\begin{abstract}
\justifying
Recent studies showed that single-machine graph processing systems can be as highly competitive as cluster-based approaches on large-scale problems. While several out-of-core graph processing systems and computation models have been proposed, the high disk I/O overhead could significantly reduce performance  in many practical cases.  In this paper, we propose GraphMP to tackle big graph analytics on a single machine. GraphMP achieves  low disk I/O overhead with three techniques. First, we design a vertex-centric sliding window (VSW) computation model to avoid reading and writing vertices on disk. Second, we propose a selective scheduling method to skip loading and processing unnecessary edge shards on disk. Third, we use a compressed edge cache mechanism to fully utilize the available memory of a machine to reduce the amount of disk accesses for edges. Extensive evaluations have shown that GraphMP could outperform existing single-machine out-of-core systems such as GraphChi, X-Stream and GridGraph by up to 30, and can be as highly competitive as distributed graph engines like Pregel+, PowerGraph and Chaos.
\end{abstract} 

}

\maketitle
\IEEEdisplaynontitleabstractindextext
\IEEEpeerreviewmaketitle

\section{Introduction}\label{sec: introduction}

\IEEEPARstart{I}{n} the era of ``Big Data'', many real-world problems, such as social network analytics and collaborative recommendation, can be represented as graph computing problems \cite{hu2014toward}. Analyzing large-scale graphs has attracted considerable interest in both  academia and industry.  However, researchers are facing significant challenges in processing big graphs, which contain billions of vertices and hundreds of billions of edges, with popular big data analysis tools like MapReduce  \cite{dean2008mapreduce} and Spark \cite{zaharia2012resilient}, since these general-purpose frameworks cannot  leverage  inherent interdependencies within graph data and common patterns of iterative graph algorithms for performance optimization \cite{mccune2015thinking}, \cite{kalavri2018high}, \cite{sun2017graphh}.

To tackle this challenge, researchers have proposed many dedicated in-memory graph processing systems over multi-core, heterogeneous and distributed infrastructures.  These systems usually adopt a vertex-centric programming model (which allows users to think like a vertex when designing parallel graph applications), and should always manage the entire input graph and all intermediate data in memory. Specifically, Ligra \cite{shun2013ligra}, Galois \cite{kulkarni2007optimistic}, GraphMat \cite{sundaram2015graphmat} and Polymer \cite{zhang2015numa} could handle generic graphs with 1-20 billion edges on a single multi-core machine. Some single-machine systems, e.g., \cite{zhong2014medusa}, \cite{khorasani2015scalable}, \cite{wang2016gunrock}, \cite{fu2014mapgraph}, \cite{zhang2015efficient}, \cite{kim2016gts}, \cite{maass2017mosaic}, \cite{nurvitadhi2014graphgen}, \cite{khorasani2014cusha}, could leverage heterogeneous devices, such as graphics processing unit (GPU), field-programmable gate array (FPGA) and Xeon Phi, to scale up graph processing performance. 

To process big graphs, which cannot be fully loaded into the memory of a single commodity machine,  three types of distributed graph engines could scale out in-memory graph processing to a cluster:
\begin{itemize}
    \item Pregel-like systems, e.g., \cite{malewicz2010pregel}, \cite{ching2015one},  \cite{yan2014pregelplus}, \cite{salihoglu2013gps}, \cite{zhou2014mocgraph}, assign each vertex and its out-going edges to a machine, and provide interaction between vertices using message passing along edges.
    \item PowerGraph \cite{gonzalez2012powergraph},  PowerLyra \cite{chen2015powerlyra} and GraphX \cite{gonzalez2014graphx} adopt the GAS (Gather-Apply-Scatter) model to improve load balance when processing power-law graphs: they split a vertex into multiple replicas, and parallelize the computation for it on different machines. 
    \item GraphPad \cite{anderson2016graphpad} and CombBLAS \cite{bulucc2011combinatorial} express common graph analyses in generalized sparse matrix-vector multiplication (SpMV) operations, and leverage high-performance computing (HPC) techniques to speed up large-scale SpMV.
\end{itemize}
However, current in-memory graph processing systems require a costly investment in powerful computing infrastructure to handle big graphs. For example, GraphX  needs more than 16TB memory to handle a 10-billion-edge graph \cite{wu2015g}.

\renewcommand\arraystretch{1.22}
\begin{table*}[]
\centering
\resizebox{1\textwidth}{!}{
\begin{threeparttable}{}
\caption{Existing approaches for large-scale graph processing.}
\label{Tab: Compare_Intro}
\begin{tabular}{@{} L{2.2cm} @{} | @{} C{1.9cm} @{} | @{} C{2.1cm} @{} | @{} C{1.65cm} @{} | @{} C{2.2cm} @{} | @{} C{2.1cm} @{} | @{} C{1.7cm} @{} | @{} C{2.5cm} @{} | @{} C{1.8cm} @{}}
\thickhline
& \multicolumn{4}{c|}{\textbf{Single Machine (CPU)}} 
& \multicolumn{2}{c|}{\textbf{Single Machine (GPU)}} 
& \multicolumn{2}{c}{\textbf{Cluster}} \\
\hline

\textbf{Data Storage} 
& \textbf{In-Memory} & \multicolumn{3}{@{}c|}{\textbf{Out-of-Core (use HDD if not indicated)}} & \textbf{In-Memory} & \textbf{Out-of-Core} 
& \textbf{In-Memory}  & \textbf{Out-of-Core} \\
\hline 

\textbf{Approaches} 
& \begin{tabular}[c]{@{}c@{}} Ligra \cite{shun2013ligra} \\ Galois\cite{kulkarni2007optimistic} \\ GraphMat \cite{sundaram2015graphmat} \\ Polymer \cite{zhang2015numa} \end{tabular} & \begin{tabular}[c]{@{}c@{}} GraphChi \cite{kyrola2012graphchi} \\ X-Stream \cite{roy2013x}  \\ VENUS \cite{cheng2015venus} \\ GridGraph \cite{zhu2015gridgraph} \end{tabular} & \begin{tabular}[c]{@{}c@{}} \textbf{GraphMP} \\ \\ \\ \\ \end{tabular}  & \begin{tabular}[c] {@{}c@{}}  (Use SSD) \\ FlashGraph \cite{da2015flashgraph} \\ TurboGraph \cite{han2013turbograph} \\ \\  \end{tabular} & \begin{tabular}[c]{@{}c@{}} Medusa \cite{zhong2014medusa} \\ Gunrock \cite{wang2016gunrock} \\ MapGraph \cite{fu2014mapgraph} \\ gGraph \cite{zhang2015efficient} \end{tabular} & \begin{tabular}[c]{@{}c@{}} (Use SSD) \\ GTS \cite{kim2016gts} \\ GGraph \cite{zhang2015efficient} \\ \\ \end{tabular} & \begin{tabular}[c]{@{}c@{}} Pregel-like:  \cite{malewicz2010pregel} \\ \cite{ching2015one} \cite{yan2014pregelplus} \cite{salihoglu2013gps} \cite{zhou2014mocgraph} \\ GAS: \cite{gonzalez2012powergraph} \cite{chen2015powerlyra} \cite{gonzalez2014graphx} \\ SpMV: \cite{anderson2016graphpad} \cite{bulucc2011combinatorial} \end{tabular} & \begin{tabular}[c]{@{}c@{}} (Use HDD) \\ GraphD \cite{yan2016efficient} \\ Chaos \cite{roy2015chaos} \\  Pregelix \cite{bu2014pregelix} \\ \end{tabular} \\ \hline 

\textbf{Scale (\#edges)} & 1-20 Billion & $\sim$100 Billion & $\sim$100 Billion & $\sim$100 Billion & 0.1-4 Billion & 4-64 Billion & 5-1000 Billion & $\sim$1 Trillion \\ \hline 

\textbf{Speed (\#edges/s)}  & 1-2 Billion & 5-100 Million & 20M-2.2B & 20-400 Million & 1-7 Billion & $\sim$0.4 Billion & 1-7 Billion & 5-200 Million \\ \hline 

\textbf{Platform Cost} & Medium & Low  & Medium & Medium & High & Medium & High & Medium \\
\thickhline
\end{tabular}
\end{threeparttable}{}
}
\end{table*}

Out-of-core systems, which maintain just a small portion of vertices and/or edges in memory, provide cost-effective solutions for big graph analytics. Single-machine engines, such as GraphChi \cite{kyrola2012graphchi},  X-Stream \cite{roy2013x}, VENUS \cite{cheng2015venus} and GridGraph \cite{zhu2015gridgraph},  break the input graph into a set of shards, each of which contains all required information to update  its associated vertices. 
In many cases, an out-of-core graph engine processes all shards in an iteration, and usually uses three stages to execute a shard: 
\begin{itemize}
  \item Loading this shard's associated vertices into memory;
  \item Processing this shard's edges from disk for updating its associated vertices;
  \item Writing updated vertices or edges back to disk.
\end{itemize}
These three stages would generate a huge amount of disk accesses, which may become performance bottleneck \cite{mccune2015thinking}. To reduce disk I/O cost, many out-of-core graph computation models have been proposed.   Representative examples include the {parallel sliding window model} (PSW) of GraphChi, the {edge-centric scatter-gather} (ESG) model of X-Stream,  the {vertex-centric streamlined processing} (VSP) model of VENUS, and the {dual sliding windows} (DSW) model of GridGraph.
These approaches try to exploit the sequential bandwidth of hard disks and to reduce the amount of disk accesses. Nonetheless, current out-of-core graph engines still have much lower  performance (5-100M edges/s) than that of in-memory graph engines (1-2B edges/s), as shown in Table \ref{Tab: Compare_Intro}. While Pregelix \cite{bu2014pregelix}, Chaos \cite{roy2015chaos} and GraphD \cite{yan2016efficient} scale out out-of-core graph processing to multiple machines, their processing performance could not be  significantly improved due to the high disk I/O cost.

In this paper, we propose GraphMP, a novel out-of-core graph processing system, to tackle big graph analytics on a single commodity machine\footnote{It is common for current commodity single machine to have more than 64GB memory. For example, a single EC2 M4 instance can have up to 256GB memory. In this work, GraphMP is deployed on a Dell R720 server with two Intel Xeon E5-2620 processors (12 cores in total), 128GB memory and 4x4TB HDDs (RAID5). } based on our previous work \cite{graphmp}. 
GraphMP employs three techniques to fully utilize available memory resources, and thus significantly reduces disk I/O overhead.
First, we design a \textbf{vertex-centric sliding window (VSW)} computation model to avoid reading and writing vertices on hard disks.
Specifically, GraphMP breaks the input graph's vertices into disjoint intervals. Each interval is associated with a shard, which contains all  edges that have destination vertex in that interval. 
During the computation, GraphMP manages all vertices of a graph in the main memory, slides a window on vertices from disks, and processes edges shard by shard.
When processing a specific shard, GraphMP first loads it into  memory, then executes user-defined functions on it to update corresponding vertices. Thus, GraphMP does not need to read or write vertices on hard disks until the end of the program, since all of them are stored in memory.
Compared to \cite{graphmp}, we enhance \textbf{selective scheduling} and \textbf{compressed edge caching} to further reduce disk I/O overhead. Specifically, with selective scheduling, inactive shards, which would not update any vertices, can be skipped to avoid unnecessary disk accesses.
Compressed edge cache mechanism could fully utilize available memory resources to cache as many as shards in memory. If a shard is cached, GraphMP would not access it from hard disks. 



\setlength{\minipagewidth}{0.485\textwidth}
\setlength{\figurewidthFour}{\minipagewidth}
\begin{figure} 
    \centering
    \begin{minipage}[t]{\minipagewidth}
    \begin{center}
    \includegraphics[width=\figurewidthFour]{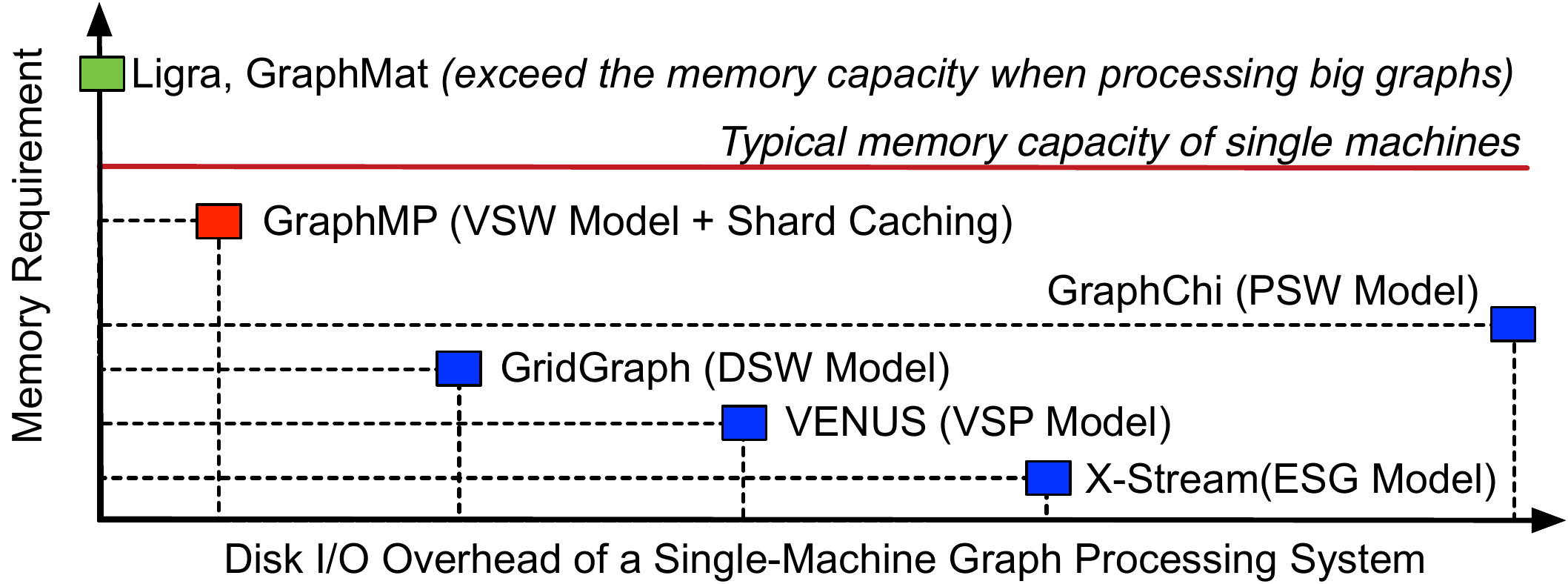}
    \end{center}
    \end{minipage}
    \centering
    \caption{Compared to  in-memory systems like Ligra, GraphMP can handle big graphs on a single machine, since it does not store all graph data in memory.  Compared to existing out-of-core systems (e.g., GraphChi, X-Stream, VENUS, GridGraph), GraphMP could fully utilize available memory of a typical server to reduce disk I/O overhead.}
\label{Fig: Intro-Want}
\end{figure}

As shown in Figure \ref{Fig: Intro-Want}, GraphMP can be distinguished from other single-machine graph engines as follows:
\begin{itemize}
  \item Compared to CPU-based in-memory approaches like GraphMat and Ligra, GraphMP does not need to manage all edges in memory, so that it can handle big graph anlytics beyond the memory limit (real-world graphs usually contain much more edges than vertices).

  \item Compared to existing HDD-based out-of-core systems like GraphChi and GridGraph, GraphMP requires more memory to manage all vertices for low disk I/O overhead. Most of the time, this is not a problem as a single commodity machine can easily fit all vertices of a big graph into memory. Take PageRank as an example, a graph with 1.1 billion vertices needs about 22.1GB
  memory to manage all vertex values. 

  \item Compared to graph engines like Gunrock and GTS, which use hetergeneous computation and storage devices (e.g., PCIe/NVMe SSD, GPU, FPGA and Xeon Phi), GraphMP is designed for big graph analytics on a commodity machine with just CPUs and HDDs.
\end{itemize}

We implement GraphMP using C++ and OpenMP, which is available at https://github.com/cap-ntu/Graphee, 
Extensive evaluations on a testbed have shown that GraphMP performs much better than existing single-machine out-of-core approaches, and  has competitive performance to popular in-memory and distributed solutions. When running PageRank, single source shortest path (SSSP) and  weakly connected components (CC) on a graph with 1.1 billion vertices, which is the largest web graph dataset could be downloaded from http://law.di.unimi.it/datasets.php, GraphMP can outperform GraphChi, X-Stream and GridGraph by a factor of up to 30.

The rest of the paper is structured as follows. In section 2, we present the system design of GraphMP. Section 3 gives quantitative comparison between our approach with other graph processing systems. The evaluation results are detailed in Section 4.  We conclude the paper in section 5.

\section{System Design}
 
In this section, we introduce the system design of GraphMP, and show how GraphMP handles big graph analytics in a single machine with vertex-centric sliding window (VSW) model, selective scheduling and compressed edge caching.

\subsection{Notations}
Graph $G=(V,E)$ contains $|V|$ vertices and $|E|$ edges. Each vertex $v$ has a unique ID $id(v)$, a value $val(v)$, an {incoming} adjacency list $\Gamma_{in}(v)$, an {outgoing} adjacency list $\Gamma_{out}(v)$, and a boolean field $active(v)$. During the computation, $val(v)$ may be updated, and $active(v)$ indicates whether $val(v)$ is updated in the last iteration. If vertex $u \in \Gamma_{in}(v)$, $u$ is an incoming neighbor of $v$, and $(u,v)$ is an in-edge of $v$. If $u \in \Gamma_{out}(v)$,  $u$ is an outgoing neighbor of $v$, and $(v,u)$ is an out-edge of $v$. $d_{in}(v)=|\Gamma_{in}(v)|$ and $d_{out}(v)=|\Gamma_{out}(v)|$ are the in-degree and out-degree of $v$, respectively. Let $val(u,v)$ denote the edge value of  $(u,v)$. In this work, if $G$ is a unweighted graph, $val(u,v)=1, \forall (u,v) \in E$. 

\setlength{\minipagewidth}{0.485\textwidth}
\setlength{\figurewidthFour}{\minipagewidth}
\begin{figure} 
    \centering
    \begin{minipage}[t]{\minipagewidth}
    \begin{center}
    \includegraphics[width=\figurewidthFour]{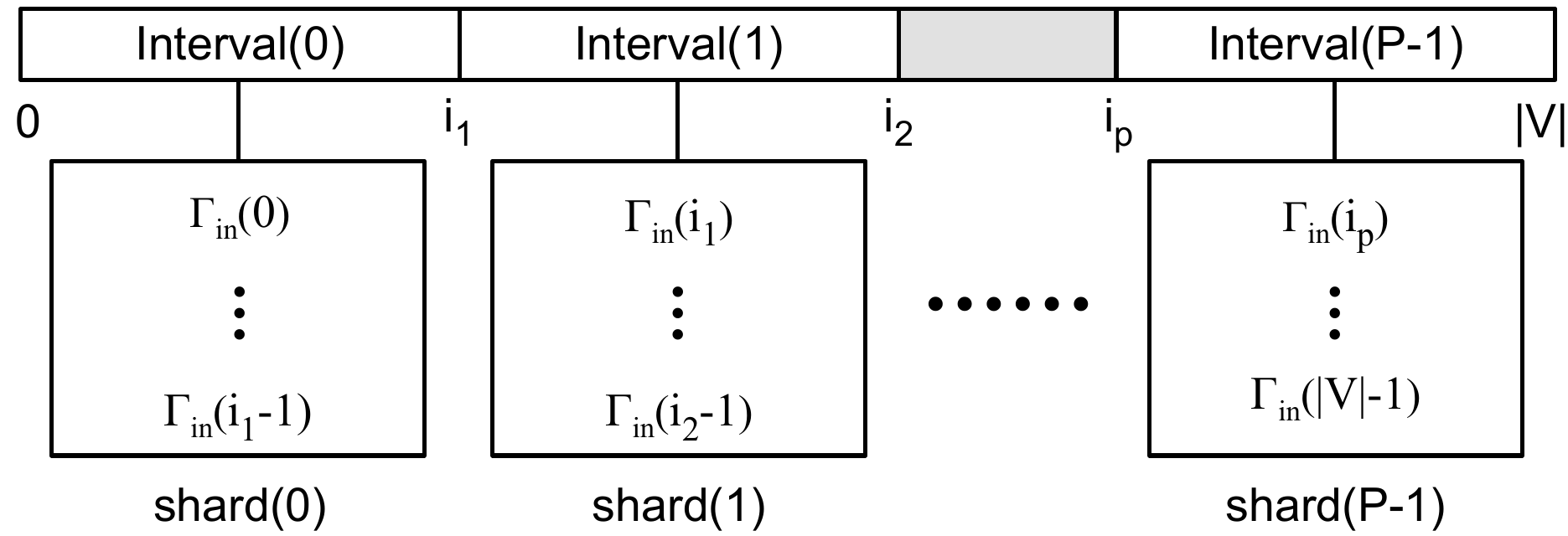}
    \end{center}
    \end{minipage}
    \centering
    \caption{The input graph's vertices are divided into $P$ intervals. Each interval is associated with a shard, which stores all edges that have destination vertex in that interval. GraphMP structures all edges of a shard in key-value pairs $(id(v), \Gamma_{in}(v))$, and stores them in the Compressed Sparse Row format.}
\label{Fig: Shard}
\end{figure}

\subsection{Graph Sharding and Data Processing}


Before vertex-centric computation, GraphMP breaks the input graph into $P$ shards in data processing stage. As shown in Figure \ref{Fig: Shard},  the vertices of $G = (V, E)$ are divided into $P$ disjoint intervals. Each interval is associated with a shard, which stores all  edges that have destination vertex in that interval.
For example, in Figure \ref{Fig: Shard}, \texttt{shard(1)} contains all edges with destination vertex $v$, where $i_1 \leq id(v) \leq i_2-1$. In this example, $i_1$ is the start vertex id of \texttt{shard(1)}, and $i_2 - 1$ is its end vertex id. Vertex intervals are chosen with two policies:
\begin{itemize}
\item Any shard can be completely loaded into memory;
\item Each shard tries to contain a similar number of edges for workload balance during computation.
\end{itemize}
In this work, each shard approximately contains 20 millions edges, so that a single shard roughly needs 80MB memory. Users can use other vertex intervals for specific applications or graph data sets.


GraphMP groups  edges in a shard by their destination, and manage them  as a sparse matrix in  Compressed Sparse Row (CSR) format. One edge is treated as a non-zero entry of the sparse matrix. The CSR format of a sparse matrix contains a \texttt{row} array, a \texttt{col} array, and a \texttt{val} array. Specifically, the \texttt{col} array stores all non-zero entries' column indices in row-major order. The \texttt{val} array contains corresponding nonzero values. 
The \texttt{row} array records each row's start point in \texttt{col} and \texttt{val} array. Figure \ref{Fig: csr} shows an example of using the CSR format to represent a 4-row sparse matrix. In this example, $\texttt{row[3]}=7$, $\texttt{row[4]}=9$. It means that the last row of the matrix contains 2 nonzero entires, whose column indices are stored in  \texttt{col[7]} and \texttt{col[8]}. The corresponding values can be accessed from \texttt{val[7]} and \texttt{val[8]}.
This work maps the edges of a shard as a sparse matrix in CSR format. For example, in Figure \ref{Fig: Shard}, \texttt{shard(1)} can be mapped as a sparse matrix with $i_2 - i_1$ rows and $|V|$ columns.
In CSR, the \texttt{col} array stores all edges' column indices in row-major order, and the \texttt{row} array records each vertex's adjacency list distribution. If the input graph is an unweighted graph, there is no need to construct the \texttt{val} array, since all edges have the same weight. In \texttt{shard(1)} of Figure \ref{Fig: Shard},  the incoming adjacency list of vertex $v$ ($i_1 \leq id(v) \leq i_2$-1)  can be accessed from: 
$$ \{ {col[row[id(v)-i_1]]}, \cdots, {col[row[id(v)+1-i_1]-1]} \}.$$


\setlength{\minipagewidth}{0.485\textwidth}
\setlength{\figurewidthFour}{\minipagewidth}
\begin{figure} 
    \centering
    \begin{minipage}[t]{\minipagewidth}
    \begin{center}
    \includegraphics[width=\figurewidthFour]{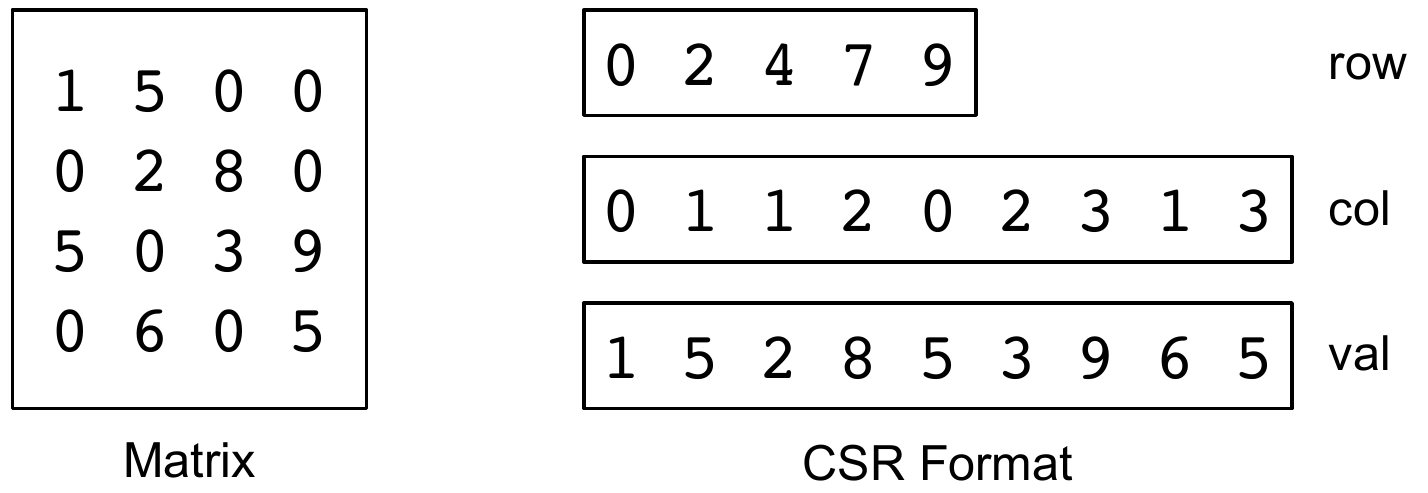}
    \end{center}
    \end{minipage}
    \centering
    \caption{An example of using the CSR format to represent a sparse matrix.}
\label{Fig: csr}
\end{figure}

In addition to edge shards, GraphMP creates two metadata files. First, a property file contains the global information of the represented graph, including the number of vertices, edges and shards, and the vertex intervals. Second, a vertex information file  stores several arrays to record the information of all vertices. It contains an array to record all vertex values (which can be the initial or updated values), an in-degree array and an out-degree array  to store each vertex's in-degree and out-degree, respectively. 


\begin{algorithm} 
\small
\caption{Compute Vertex Intervals}\label{Alg: vertex2itervals}
shard\_id = 0, vertex\_id = 0, edge\_num = 0 \\
shard[shard\_id].start\_vertex\_id = 0 \\
\While {vertex\_id $<$ $|V|$} {
  edge\_num += $\Gamma_{in}(V_{\text{vertex\_id}})$ \\
  \If {edge\_num $>$ threshold\_edge\_num} {
    shard[shard\_id].end\_vertex\_id = vertex\_id - 1 \\
    shard\_id += 1 \\
    shard[shard\_id].start\_vertex\_id = vertex\_id \\
    edge\_num = $\Gamma_{in}(V_{\text{vertex\_id}})$ \\
  }
  vertex\_id += 1
}
shard[shard\_id].end\_vertex\_id = $|V|$ - 1 \\
\end{algorithm}

GraphMP uses  three steps to preprocess a graph.
In the first step, GraphMP scans the graph to record each vertex's in-degree. With this information, GraphMP computes each shard's associated vertex interval based on Algorithm \ref{Alg: vertex2itervals}. In this method, \emph{threshold\_edge\_num} denotes the max number of edges a shard could contain, which is user defined and should be no greater than the graph's max in-degree. With this simple method, GraphMP quickly divides input vertices into a set of intervals, and each shard is allocated with a start vertex id and an end vertex id. This method also guarantees that  each shard is small enough to be loaded into memory, and tries to let each shard contain a similar number of edges. 
In the second step, GraphMP sequentially reads graph edges from disk, and appends each edge to a shard file based on its destination and computed vertex intervals. 
In the third step, GraphMP transforms all shard files to the CSR format, and persists them on disk. After these three steps, all edges are actually sorted and grouped by their destination vertices.

After the data preprocessing, GraphMP is ready to perform vertex-centric computation based on the VSW model. GraphMP only needs to perform data preprocessing one time, then could execute any graph applications using the same partitioned data. As a comparison, GraphChi needs to preprocess  input graph again before executing a new type of a graph application. For example, GraphChi cannot use the same partitioned graph for PageRank to run SSSP \cite{kyrola2012graphchi}.

\setlength{\minipagewidth}{0.485\textwidth}
\setlength{\figurewidthFour}{\minipagewidth}
\begin{figure} 
    \centering
    \begin{minipage}[t]{\minipagewidth}
    \begin{center}
    \includegraphics[width=\figurewidthFour]{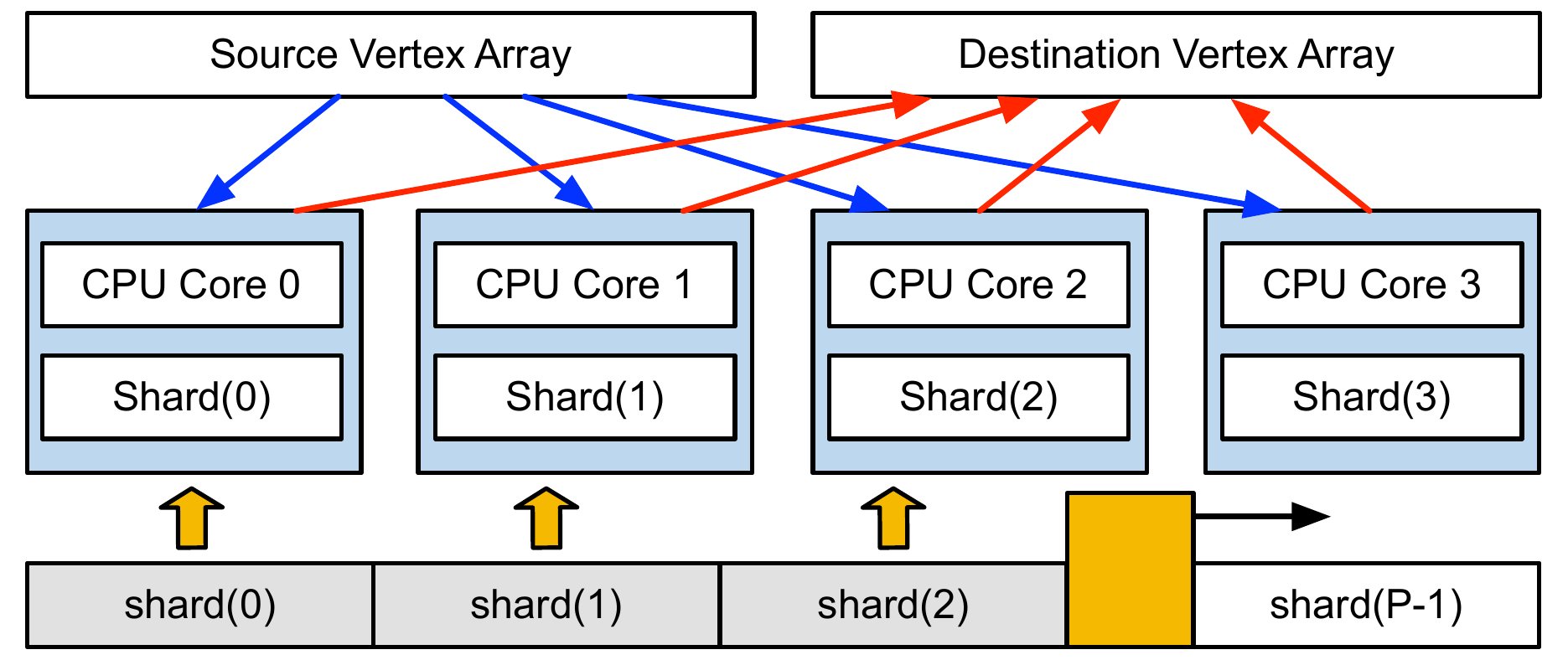}
    \end{center}
    \end{minipage}
    \centering
    \caption{The VSW computation model. GraphMP slides a window on vertices, and makes each CPU core process a shard at a time. When processing a shard, a CPU core continually pulls required vertex values from memory, and pushes updated ones to another array in memory.}
\label{Fig: Sliding}
\end{figure}

\subsection{Vertex-Centric Sliding Window Computation}


GraphMP slides a window on vertices, and processes edges shard by shard on a single machine with $N$ CPU cores, as shown in Figure \ref{Fig: Sliding} and Algorithm \ref{Alg: VSW}.
During the computation, GraphMP maintains two vertex arrays in  memory until the end of the program: \texttt{SrcVertexArray} and \texttt{DstVertexArray}. The \texttt{SrcVertexArray} stores latest vertex values, which are the input of the current iteration. Updated vertex values are written into the \texttt{DstVertexArray}, which are used as the input of the next iteration.
GraphMP  uses OpenMP to parallelize the computation (line 3 of Algorithm \ref{Alg: VSW}): each CPU core processes a shard at a time.
When processing a specific shard, GraphMP first loads it into memory (line 6), then executes user-defined vertex-centric functions, and writes the results to the \texttt{DstVertexArray} (line 7-8). 
Given a vertex, if its value is updated, we call it an active vertex. Otherwise, it is inactive.
After processing all shards, GraphMP records all active vertices in a list (line 9).  This list could help GraphMP to avoid loading and processing inactive shards in the next iteration (line 5),  which would not generate any updates (detailed in Section 2.4). The  values of \texttt{DstVertexArray} are used as the input of next iteration (line 10).
The program terminates if it does not generate any active vertices (line 2).

\begin{algorithm} 
\small
\caption{Vertex-Centric Sliding Window Model}\label{Alg: VSW}
init (src\_vertex\_array, dst\_vertex\_array) \\ 
\While  {active\_ratio $>$ 0} {
  \# pragma omp parallel for num\_threads(N)\\
  \For {shard $\in$ all\_shards} { 
    \If {active\_vertex\_ratio $>$ threshold\_active\_ratio  \textbf{or} Bloom\_filter[shard.id].has(active\_vertices)} {
    {load\_to\_memory(shard)} \\
      \For {v $\in$ shard.associated\_vertices} {
        {dst\_vertex\_array[v.id] $\gets$ \textbf{update}(v, src\_vertex\_array)} \\
      }
      }
    }
    active\_vertices = \{vertices with updated values\} \\
    src\_vertex\_array $\gets$ dst\_vertex\_array \\
    active\_ratio $\gets$  $|$active\_vertices$|$ $/$ vertex\_num 
}
\end{algorithm}


Users need to define two functions for a particular application: \texttt{Init} and \texttt{Update}. 
Specifically, the \texttt{Init} function takes \texttt{SrcVertexArray} and \texttt{DstVertexArray} as inputs,
$$\texttt{Init(SrcVertexArray, DstVertexArray)},$$
and initialize the values of all vertices. 
The \texttt{Update} function accepts a vertex and  \texttt{SrcVertexArray} as inputs, 
$$\texttt{Update(InputVertex, SrcVertexArray)},$$
and should return two results: an updated vertex value which should be stored in \texttt{DstVertexArray}, and a boolean value to indicate whether the input vertex updates its value. Specifically, this function allows the input vertex to pull the values of its incoming neighbors from  \texttt{SrcVertexArray} along the in-edges, and uses them to update its value.

\SetKwProg{Fn}{Function}{}{}
\begin{algorithm}
\small
\caption{PageRank, SSSP and CC in GraphMP}\label{Alg: GAB-PR}
// \emph{vertex\_value (Double) is the rank value}\\
\Fn{PR\_Init(Array src\_vertex, Array dst\_vertex)}{
    \For {i $\in$ range(num\_vertex)} {
        src\_vertex[i] = dst\_vertex[i] = 1 / num\_vertex \\
    }
    active\_vertices = \{all vertices\}
}
\Fn{PR\_Update(Vertex v, Array src\_vertex)}{
    sum = 0 \\
    \For {e $\in$ v.incoming\_neighbours} {
        sum += src\_vertex[e.source] / e.source.out\_deg \\
    }
    updated\_value = 0.15 / num\_vertex + 0.85 * sum \\
    \Return updated\_value\\
}
\nonl  \hrulefill \\
// \emph{vertex\_value (Long) is the distance to the source vertex}\\
\Fn{SSSP\_Init(Array src\_vertex, Array dst\_vertex)}{
    \For {i $\in$ range(num\_vertex)} {
        \If {i == source\_vertex.id} {
            src\_vertex[i] = dst\_vertex[i] = 0 \\
        }
        \Else {
            src\_vertex[i] = dst\_vertex[i] = $\infty$ \\
        }
    }
    active\_vertices = \{source\_vertex\}
}
\Fn{SSSP\_Update(Vertex v, Array src\_vertex)}{
    d = $\infty$ \\
    \For {e $\in$ v.incoming\_neighbours} {
        d = min (src\_vertex[e.source] + (e,u).val, d)  \\
    }
    updated\_value = min (d, v.value) \\
    \Return updated\_value \\
}
\nonl  \hrulefill \\
// \emph{vertex\_value ($Long$) is the vertex group id}\\
\Fn{CC\_Init(Array src\_vertex, Array dst\_vertex)}{
    \For {i $\in$ range(num\_vertex)} {
        src\_vertex[i] = dst\_vertex[i] = i \\
    }
    active\_vertices = \{all vertices\}
}
\Fn{CC\_Update(Vertex v, Array src\_vertex)}{
    subgraph\_id = $\infty$ \\
    \For {e $\in$ v.incoming\_neighbours} {
        subgraph\_id = min (src\_vertex[e.source], subgraph\_id)  \\

    }
    updated\_value = min (subgraph\_id, v.value) \\
    \Return updated\_value  \\
}
\end{algorithm}

We implement three popular graph applications (PageRank, SSSP and CC) using \texttt{Init} and \texttt{Update} in Algorithm \ref{Alg: GAB-PR}. 
PageRank is an algorithm used to  measure the importance of website pages. SSSP is  used to find shortest paths from a source vertex to all other vertices in the graph. CC can detect whether any two vertices of the graph are connected to each other by paths. In PageRank, the vertex value type is \texttt{Double} to store the rank value of a vertex. The graph is initialized before the first iteration, and the value of each vertex is $1/vertex\_num$ (line 3-4). All vertices are set to be active in the initialization phase (line 5).  During the iterative computation, each vertex accumulates vertex values along its in-edges into $sum$ (line 8-9), and sets its own rank value to $0.15/vertex\_num+0.85*sum$ (line 10). The two hyper-parameters "$0.15$" and "$0.85$" are adopted from Google \cite{malewicz2010pregel}, which help the algorithms converges smoothly.
In SSSP, the vertex value type is \texttt{Long} to store the minimum distance from the source vertex (for example, vertex 0). Before the first iteration, the source vertex initializes its value to zero, and other vertex values are initialized to $\infty$ (line 14-18). Only the source vertex is set to be active in the initialization phase (line 19). During the computation, each vertex connects its neighbor vertices along in-edges (line 22-23), and tries to find a shorter path to the source vertex (line 24). 
When running CC on undirected graphs, the vertex value type is \texttt{Long} to store the subgraph ID. If two vertices have the same subgraph ID, they are connected to each other by paths. Before the first iteration, the value of each vertex is initialized to its vertex ID (line 28-29).  All vertices are set to be active in the initialization phase (line 30).  During the computation, each vertex checks the subgraph ID of its neighbors, and  overwrites its own subgraph ID with the max vertex ID received from its neighbors. This continues until convergence (line 33-35).

When using multiple CPU cores to process graph shards in parallel, GraphMP does not require locks or atomic operations. This property could  improve the graph  processing performance considerably. As shown in Figure \ref{Fig: Sliding} and Algorithm \ref{Alg: VSW} (line 3), in each iteration, GraphMP uses one CPU core to process a shard for updating its associated vertices, and could process $N$ shards in parallel when having $N$ GPU cores. Given a vertex $v$, \texttt{SrcVertexArray[v.id]} may be accessed as input by multiple CPU cores at the same time. Due to GraphMP's graph sharding strategy (specifically all in-edges of a vertex are managed in the same shard), \texttt{DstVertexArray[v.id]} is computed from edges located in a single shard, and could be written by a single CPU core in each iteration. Therefore, there is no need to use locks or  atomic operations to avoid data inconsistency issues on \texttt{SrcVertexArray} and \texttt{DstVertexArray}. As a comparison, GridGraph should use an atomic operation to process each edge, since multiple cores may simultaneously  update the same vertex \cite{zhu2015gridgraph}.


In Figure \ref{Fig: Example_Graph}, we show an example of how GraphMP runs PageRank.  The input graph is partitioned into three shards, each of which contains two vertices and their adjacency lists.  At the beginning of PageRank, all  vertex values are initiated to be $1/num\_vertex = 0.14$. GraphMP slides a window on vertices, and lets each CPU core process a shard at a time. When processing shard 0 on a CPU core, GraphMP pulls the values of vertex 1, 3 from \texttt{SrcVertexArray}, then use them to compute the updated value for vertex 0, and writes it to \texttt{DstVertexArray[0]}.  
After processing all 3 shards, GraphMP uses the values of \texttt{DstVertexArray} to replace the values of \texttt{SrcVertexArray}, and starts the next iteration if there are any active vertices.

\setlength{\minipagewidth}{0.485\textwidth}
\setlength{\figurewidthFour}{\minipagewidth}
\begin{figure} 
    \centering
    \begin{minipage}[t]{\minipagewidth}
    \begin{center}
    \includegraphics[width=\figurewidthFour]{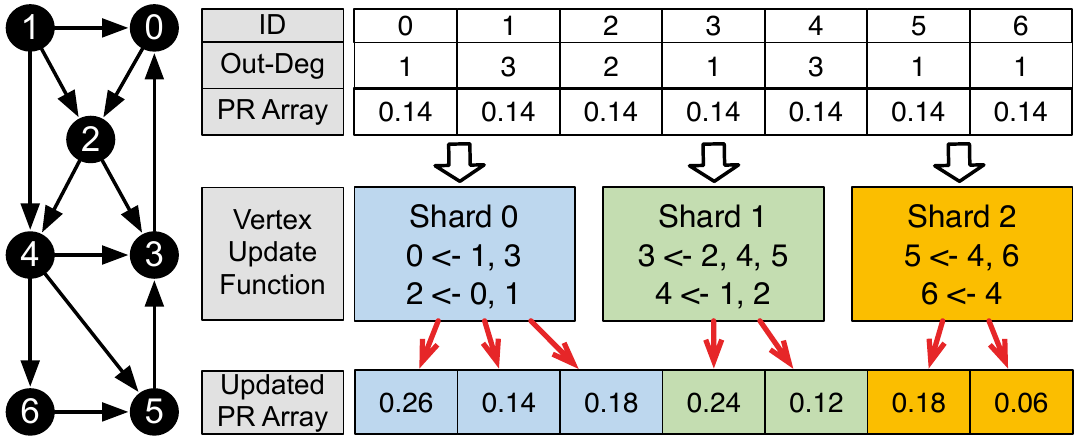}
    \end{center}
    \end{minipage}
    \centering
    \caption{An example of the first iteration of PageRank on GraphMP.}
\label{Fig: Example_Graph}
\end{figure}

\subsection{System Optimizations}

GraphMP employs two optimization techniques (specifically selective scheduling and compressed edge caching) to further reduce the disk I/O overhead and improve the graph processing performance. 

\subsubsection{Selective Scheduling}

For many graph applications, such as PageRank, SSSP and CC, a lot of vertices converge quickly and would not update their values in the rest iterations. Given a shard, if all source vertices of its associated edges are inactive,  it is an inactive shard. An inactive shard would not generate any updates in  the following iteration.
Therefore, it is unnecessary to load and process these inactive shards.

To solve this problem, we leverage Bloom filters to detect inactive shards, so that GraphMP could avoid unnecessary disk accesses and processing. A Bloom filter is a memory-efficient data structure, which can rapidly test whether an element is a member of a set by using multiple hash functions.  GraphMP manages a Bloom filter for each shard to record the source vertices of its edges. When processing a shard, GraphMP uses its Bloom filter to check whether it contains any active vertices. If yes, GraphMP would continue to load and process the shard. Otherwise, GraphMP would skip it. For example, in Figure \ref{Fig: Example_Graph}, when the sliding window is moved to shard 2, its Bloom filter could tell GraphMP whether vertex 4, 6 have changed their values in the last iteration. If there are no active vertices, the sliding window would skip shard 2, since it cannot not update vertex 5 or 6 after the processing.

GraphMP only enables selective scheduling when the ratio of active vertices is lower than a threshold. If the active vertex ratio is high, nearly all shards contain at least one active vertex. 
In this case, GraphMP wastes a lot of time on detecting inactive shards, and would not reduce any unnecessary disk accesses. 
As shown in Algorithm \ref{Alg: VSW} Line 5, GraphMP starts to detect inactive shards when the active vertex ratio is lower than a threshold. 
In this paper, we use $0.001$ as the threshold  according to our experiment data. Users can choose a better value for specific applications.

\subsubsection{Compressed Edge Caching}

We design a cache system in GraphMP to reduce the amount of disk accesses for edges.
The VSW computation model requires  storing all vertices and shards under processing in the main memory. 
These data would not consume all available memory resources of a single machine. 
For example, given a server with 24 CPU cores and 128GB memory, when running PageRank on a graph with 1.1 billion vertices, GraphMP uses 21GB memory to store all data, including \texttt{SrcVertexArray}, \texttt{DstVertexArray}, the  out-degree array, Bloom filters, and the shards under processing.
It motivates us to build an in-application cache system to fully utilize available memory to reduce the disk I/O overhead.
Specifically, when GraphMP needs to process a shard, it first searches the cache system. If  there is a cache hit, GraphMP can process the shard without disk accesses. Otherwise, GraphMP loads the target shard from  disk, and leaves it in the cache system if the cache system is not full.  

\renewcommand\arraystretch{1}
\begin{table}
\centering
\caption{Compression ratio and processing throughput per CPU core.}
\label{Tab: compress1}
\resizebox{0.49\textwidth}{!}{
\begin{tabular}{@{}lcccccc@{}}
\thickhline
& \multicolumn{3}{c}{\textbf{Compression Ratio}} & \multicolumn{3}{c}{\textbf{Throughput (MB/s)}} \\
& snappy  & zlib-1  & zlib-3  & snappy & zlib-1  & zlib-3           \\  \hline
\textbf{Twitter} & 1.75 & 2.78 & 3.22  & 870 & 55  & 46    \\ 
\textbf{UK-2007} & 1.89 & 3.71 & 4.54 & 947  & 58  & 53 \\ 
\textbf{UK-2014} & 1.96 & 4.34  & 5.26 & 903   & 65 & 50 \\ 
\textbf{EU-2015} & 1.96 & 4.35  & 5.88  & 890 & 62  & 56 \\ \thickhline
\end{tabular}
}
\subcaption{}
\resizebox{0.49\textwidth}{!}{
\begin{tabular}{@{}l@{}C{1.45cm}@{}C{1.45cm}@{}C{1.45cm}@{}C{1.45cm}@{}C{1.45cm}@{}}
\thickhline
 & \textbf{\begin{tabular}[c]{@{}c@{}}Size (GB)\\ (CSV)\end{tabular}} & \textbf{\begin{tabular}[c]{@{}c@{}}Size (GB) \\ (raw)\end{tabular}} & \textbf{\begin{tabular}[c]{@{}c@{}}Size (GB)\\ (snappy)\end{tabular}} & \textbf{\begin{tabular}[c]{@{}c@{}}Size (GB)\\ (zlib-1)\end{tabular}} & \textbf{\begin{tabular}[c]{@{}c@{}}Size (GB)\\ (zlib-3)\end{tabular}} \\ \hline 
\textbf{Twitter} & 24 & 6.5 & 3.7 & 2.3 & 2 \\ 
\textbf{UK-2007} & 94 & 23 & 12 & 6.2 & 5 \\ 
\textbf{UK-2014} & 874 & 196 & 100 & 45 & 37 \\ 
\textbf{EU-2015} & 1700 & 362 & 185 & 80 & 62 \\ \thickhline
\end{tabular}
}
\end{table}

GraphMP can compress cached shards to improve the amount of cached edge shards and further reduce disk I/O cost.
Table \ref{Tab: compress1} shows that popular compressors can efficiently reduce the size of graph datasets. We use four  real-world graphs as inputs: Twitter, UK-2007, UK-2014 and EU-2015. Section \ref{sec: performance} gives more detail about these four graph datasets.  We see that zlib-3 could compress EU-2015 by a factor of $5.88$. While GraphPS needs additional decompression time, the edge cache system still provides higher throughput than hard disks. For example, snappy can decompress an edge shard at a rate of 903MB/s using a single CPU core. In contrast, we can only achieve up to $310$MB/s sequential disk read speed with RAID5, and the available disk bandwidth is shared by all CPU cores.

In this work, we use two compressors (snappy and zlib), and consider 5 cache modes:
\begin{itemize}
\item Cache-0: Use system page cache without edge cache.
\item Cache-1: Cache uncompressed  edge shards.
\item Cache-2: Cache  shards compressed by snappy.
\item Cache-3: Cache  shards compressed by zlib-1.
\item Cache-4: Cache  shards compressed by  zlib-3.
\end{itemize}
GraphMP can automatically select the most suitable cache mode,  considering disk I/O and decompression cost. When having limited memory, it is crucial to select compressors with high compression ratio for low disk I/O overhead. If the memory is large,  caching shards with low compression ratio can reduce decompression overhead without increasing disk I/O overhead. Let $C$ denote the memory size of the cache system,  $S$ is the input graph's size, and $\gamma_i$ is the estimated compression ratio of  cache mode-$i$.  GraphMP selects minimal $i$ constrained by $S/\gamma_i  \leq C$. If no mode satisfies this constraint, GraphMP uses mode-4 with highest compression ratio. In this case, GraphMP caches as many shards as possible in memory, and reads other shards from disk during computation.
In this work, $\gamma_0 = 1, \gamma_1 = 2, \gamma_2 = 4, \gamma_3 = 5$, according to Table \ref{Tab: compress1}.

\section{Theoretical Comparison}

We compare our proposed VSW model with four popular graph computation models: the {parallel sliding window model} (PSW) of GraphChi, the {edge-centric scatter-gather} (ESG) model of X-Stream,  the {vertex-centric streamlined processing} (VSP) model of VENUS and the {dual sliding windows} (DSW) model of GridGraph. All systems partition the input graph into $P$ shards or blocks, and run applications using $N$ CPU cores. Let $C$ denote the size of a vertex record, and $D$ is the size of one edge record. 
For fair comparison, we assume that the neighbors of a vertex are randomly chosen, and the average degree is $d_{avg} = |E|/|V|$. We disable selective scheduling, so that all system should process all edges in each iteration. We use the amount of data read and write on disk per iteration, and the memory usage as the evaluation criteria.
Table \ref{Tab: Model Compare} summarizes the analysis results.

\renewcommand\arraystretch{1}
\begin{table*}[]
\centering
\caption{Analysis of graph computation models. $C$ is the size of a vertex value, $D$ is the size of an edge value, $P$ is the number of partitioned shards or blocks of a graph, $d_{avg}$ denotes the graph's average degree, $\delta \approx (1 - e^{-d_{avg}/P})P$, $\theta$ denotes GraphMP's cache hit ratio and $0 \leq \theta \leq 1$.}
\label{Tab: Model Compare}
\begin{tabular}{@{} l @{} >{\centering}m{3.3cm} @{} >{\centering}m{2.9cm} @{} >{\centering}m{2.9cm} @{} >{\centering}m{2.9cm} @{} c @{}}
\thickhline
\textbf{Category} & \textbf{PSW (GraphChi)} & \textbf{ESG (X-Stream)} & \textbf{VSP (VENUS)} & \textbf{DSW (GridGraph)} & \textbf{VSW (GraphMP)} \\ \hline
\textbf{Data Read} &  $C|V|+2(C+D)|E|$ & $C|V| + (C+D)|E|$ & $C(1 + \delta)|V| + D|E|$ & $\;\; C\sqrt{P}|V| + D|E| \;\;$  & $\theta D|E|$  \\ 
\textbf{Data Write} &  $C|V|+2(C+D)|E|$  &  $C|V| + C|E|$ &  $C|V|$  & $C\sqrt{P}|V|$ & $0$  \\ 
\textbf{Memory Usage} &  $(C|V|+2(C+D)|E|)/P$  &  $C|V|/P$ &  $C(2 + \delta)|V|/P$  &  $2C|V|/\sqrt{P}$ & $\;\; 2C|V| + ND|E|/P$ \;\; \\ 
\textbf{Preprocessing I/O Cost \;\;} &  $(C+5D)|E|$  &  $2D|E|$ & $4D|E|$  &  $6D|E|$ & $5D|E|$  \\ \thickhline
\end{tabular}
\end{table*}

\subsection{{The PSW Model of GraphChi}}

Under PSW, GraphChi splits the vertices $V$ of graph $G = (V, E)$ into $P$ disjoint intervals. For each interval, GraphChi associates a shard, which stores all the edges that have destination in the interval. Edges are stored in the order of their source. 
Unlike GraphMP where each vertex can access the values of its neighbors from  \texttt{SrcVertexArray}, GraphChi accesses such values from the edges. Thus, the data size of each edge in GraphChi is $(C+D)$ \cite{cheng2015venus}. In addition, GraphChi stores updated vertex values in a single file as flat array of user-defined type. For each iteration, GraphChi uses three steps to process a shard: (1) loading its associated vertices, in-edges and out-edges from disk into memory; (2) updating the vertex values; and (3) writing updated vertices and edges to disk. In step (1), GraphChi loads each vertex (which incurs $C|V|$ data read), and accesses in-edges and out-edges of each vertex (which incurs $2(C+D)|E|$ data read). In step (3), GraphChi writes updated vertices into disk (which incurs $C|V|$ data write), and writes each edge twice (which incurs $2(C+D)|E|$ data write) if the computation updates edges in both directions. With the PSW model, the data read and write in total are both  $C|V|+2(C+D)|E|$. In step (2), GraphChi needs to keep $|V|/P$ vertices and their in-edges, out-edges in memory for computation. The memory usage is $(C|V|+2(C+D)|E|)/P$.

GraphChi uses 3 steps to divide a graph into $P$ shards: (1)  counting the in-degree of each vertex (which incurs $D|E|$ data read) and dividing vertices into $P$ intervals , 
(2) writing each edge to a temporary scratch file of the owning shard (which incurs $D|E|$ data read and $D|E|$ data write); and (3) sorting edges and writing each file in compact format (which incurs $D|E|$ data read and $(C+D)|E|$  data write).  The total I/O cost of the preprocessing is $(C+5D)|E|$.

\subsection{{The ESG Model of X-Stream}}

X-Stream splits  the input graph's vertices into $P$ partitions, each of which could fits in high-speed memory. Furthermore, X-Stream assigns edges to $P$ partitions, such that the edge list of a partition consists of all edges whose source vertex is in the partition’s vertex set. Then, X-Steam processes the graph one partition at a time  with two phases under the ESG model. In phase (1), when processing a graph partition, X-Stream first loads its associated vertices into memory, and processes its out-edges in a streaming fashion: generating and propagating updates (the size of an update is $C$) to corresponding values on disk. In this phase, the size of data read is $C|V| + D|E|$, and the size of data write is $C|E|$.  In  phase (2), X-Stream processes all updates and uses them to update vertex values on disk. In this phase, the size of data read is $C|E|$, and the size of data write is $C|V|$. With the ESG model, the data read and write in total are $C|V| + (C+D)|E|$ and $C|V| + C|E|$, respectively. X-Stream only needs to keep the vertices of a partition in memory, so the memory usage is $C|V|/P$.

X-Stream needs one step for data preprocessing. Specifically, before the computation, it reads edges from disks in sequence, and append them to corresponding files on disks. X-Stream does not need to sort edge lists or convert the data structure during preprocessing. Thus, the I/O cost of the preprocessing is $2D|E|$.

\subsection{{The VSP Model of VENUS}}

VENUS evenly splits $|V|$ vertices into $P$ disjoint intervals. Each interval is associated with a g-shard (which stores all edges with destination in that interval), and a v-shard (which contains all vertices appear in that g-shard). For each iteration, VENUS processes g-shards and v-shards sequentially in three steps: (1) loading a v-shard into the main memory, (2) processing its corresponding g-shard in a streaming fashion,  (3) writing updated vertices to disk. In step (1), VENUS needs to process all edges once, which incurs $D|E|$ data read. In step (3), all updated vertices are written to disk, so the data write is $C|V|$.  According to Theorem 2 in \cite{yan2015effective}, each vertex interval contains $|V|/P$ vertices, and each v-shard contains up to $|V|/P + (1-e^{-d_{avg}/P})|V|$ entries. Therefore, the data read and write are $C(1 + \delta)|V| + D|E|$ and $C|V|$ respectively, where  $\delta \approx (1 - e^{-d_{avg}/P})P$. VENUS needs to keep a v-shard and its updated vertices in memory, so the memory usage is $C(2 + \delta)|V|/P$.

VENUS uses two steps for preprocessing. Since VENUS evenly splits the set of vertices into $P$ disjoint intervals, there is no need to count the degree of each vertex first. Therefore, VENUS reads the input graph data sequentially, adds each encountered edge into a buffer according to its destination, and writes the sorted edges into an intermediate file when an buffer is full. The second step performs a $k$-way merge on all intermediate files resulted from the first step to construct required data structure. Thus, edges are grouped by their destination. The I/O cost of the preprocessing is $4D|E|$.

\subsection{{The DSW Model of GridGraph}}

GridGraph groups the input graph's $|E|$ edges into a ``grid'' representation. Specifically, the $|V|$ vertices are divided into $\sqrt{P}$ equalized vertex chunks and $|E|$ edges are partitioned into $\sqrt{P} \times \sqrt{P}$ blocks according to the source and destination vertices.  Each edge is placed into a block using the following rule: the source vertex determines the row of the block, and the destination vertex determines the column of the block. GridGraph processes edges block by block. 
GridGraph uses 3 steps to process a block in the $i$-th row and $j$-th column: (1) loading the $i$-th source vertex chunk  and  the $j$-th destination vertex chunk into memory; (2) processing  edges in a streaming fashion for updating the destination vertices; and (3) writing the destination vertex chunk to disk if it is not required by the next block. 
After processing a column of blocks, GridGraph reads $|E|/\sqrt{P}$ edges and $|V|$  vertices, and writes $|V|/\sqrt{P}$ vertices to disk.  The data read and write are $C\sqrt{P}|V| + D|E|$ and $C\sqrt{P}|V|$, respectively. During the computation, GridGraph  needs to keep two vertex chunks in memory, so the memory usage is $2C|V|/\sqrt{P}$.

GridGraph needs three steps for data processing based on the provided program. In the first step, GridGraph reads edges sequentially, calculates the block that an edge  belongs to, and appends the edge to the corresponding block file. To improve I/O throughput, GridGraph combines the generated $\sqrt{P} \times \sqrt{P}$ block files into a column-oriented file and a row-oriented file in step 2 and step 3. The I/O cost of the preprocessing is $6D|E|$.

\subsection{{The VSW Model of GraphMP}}

GraphMP keeps all source  and destination vertices in the main memory during the vertex-centric computation. Therefore, GraphMP would not incur any disk write for vertices in each iteration until the end of the program. In each iteration, GraphMP should use $N$ CPU cores to process $P$ edge shards in parallel, which incurs $D|E|$ data read. Since GraphMP uses a compressed edge cache mechanism, the actual size of data read of GraphMP is $\theta D|E|$, where $0 \leq \theta \leq 1$ is the cache miss ratio. During the computation, GraphMP manages $|V|$ source vertices (which are the input of the current iteration) and  $|V|$ destination vertices (which are the output the current iteration and the input of the next iteration) in memory, and each CPU core loads $|E|/P$ edges in memory. The total memory usage is $2C|V| + ND|E|/P$. 
As discussed in Section 2, GraphMP needs three steps for data preprocessing, and its I/O cost is $5D|E|$.

\subsection{Discussion}

PSW, ESG, VSP, DSW and VSW adopt a similar way to process large-scale graphs on a single machine: they partition a big graph into small shards or blocks, and uses limited computation resources to sequentially process these small graph shards. As detailed in this section, each graph computation model has its own graph sharding policy, data structure and vertex-centric computation flow, which could significantly affect the I/O overhead of graph applications. As shown in Table \ref{Tab: Model Compare}, the VSW model of GraphMP could reduce the amount of data reads and writes on disks than other models at the cost of higher memory usage. More specifically, VSW manages all vertices in memory and only needs to access edges from hard disks during the computation. With the help of compressed edge caching, VSW further reduces the amount of data reads by caching a portion of edge shards in memory. As a comparison, PSW, ESG, VSP and DSW should read both vertices and edges from disks at each iteration. Additionally, VSW could directly update vertex values in memory without writing those data to disks. As a compression, VSP and DSW should frequently write updated vertices to disks. PSW and ESG  need to update on-disk edges since they use a single data structure to manage both edges and vertices.
In Section IV, we use experiments to show that  a single commodity machine could provide sufficient memory for processing big graphs with the VSW model. Also, GraphMP has similar data preprocessing I/O cost with other graph engines.

\section{Performance Evaluations}\label{sec: performance}

We evaluate the performance of GraphMP using a Dell R720 server with three applications (PageRank, SSSP, CC) and four directed graph  datasets. The physical machine contains two Intel Xeon E5-2620 CPUs, 128GB memory, 4x4TB HDDs (RAID5). Table \ref{Tab: Datasets} shows the information of used datasets: Twitter, UK-2007, UK-2014 and EU-2015. Twitter is a social network graph crawled in 2010 \cite{kwak2010twitter}, showing connections between twitter users. UK-2007 and UK-2014 are two web graphs crawled in 2007 and 2014 respectively, showing links between web pages in the .uk domain. EU-2015 is a web graph crawled in 2015, showing page links in European Union countries. EU-2015 is our largest graph dataset, containing 1.1 billion vertices and 91.8 billion edges. If we store the raw graph in the CSV format, EU-2015 is a 1.7TB file. All datasets are real-word power-law graphs. As shown in Figure \ref{Fig: Input_Data}, in all four graphs, most vertices have relatively few neighbors while a few have many neighbors. Since we run CC on undirected graphs, we need to convert the input directed graphs into undirected graphs, and use undirected graphs as the input of CC.

\setlength{\minipagewidth}{0.235\textwidth}
\setlength{\figurewidthFour}{\minipagewidth}
\begin{figure} 
    \centering
    \begin{minipage}[t]{\minipagewidth}
    \begin{center}
    \includegraphics[width=\figurewidthFour]{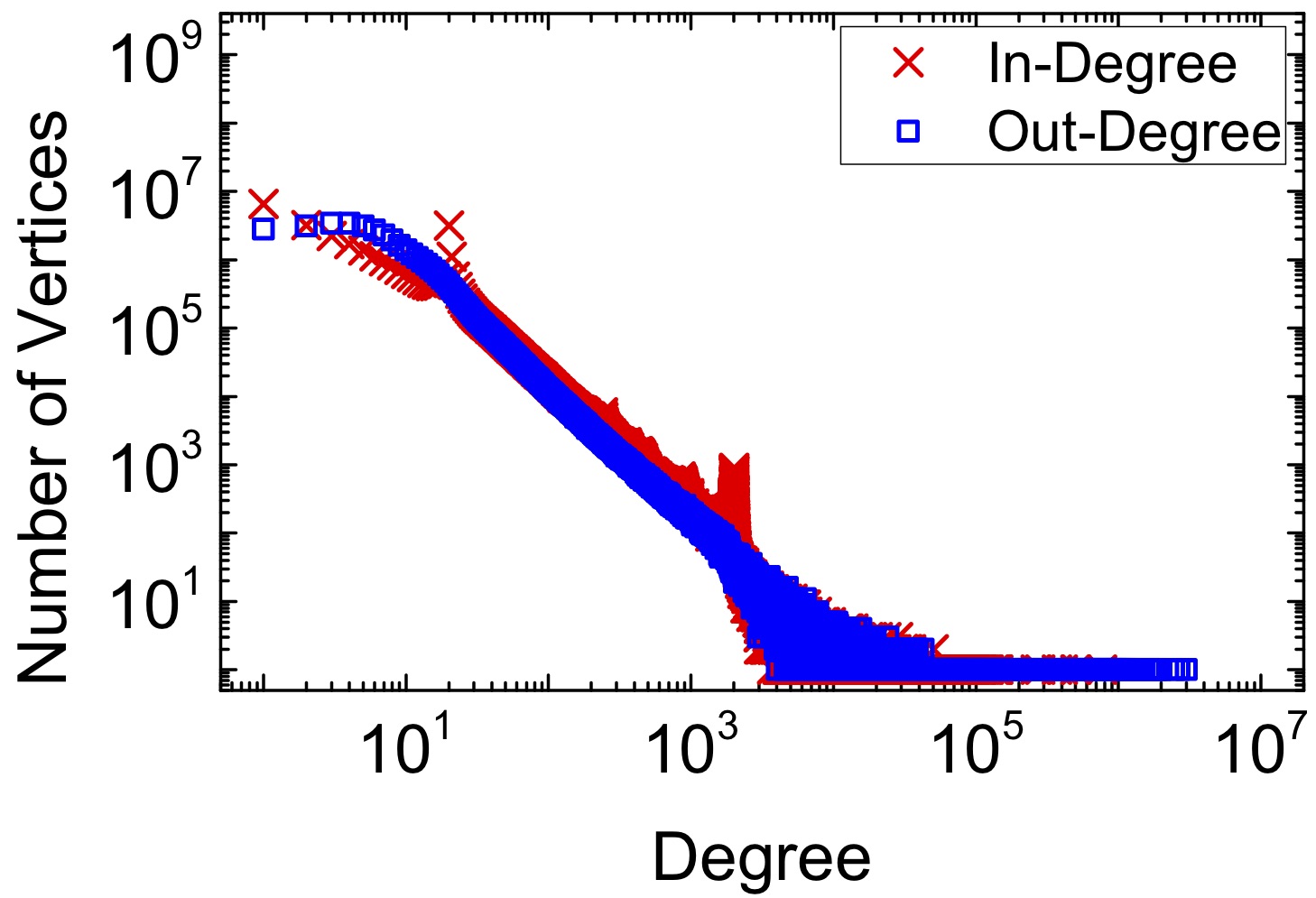}
    \subcaption{(a) Twitter}
    \end{center}
    \end{minipage}
    \centering
    \begin{minipage}[t]{\minipagewidth}
    \begin{center}
    \includegraphics[width=\figurewidthFour]{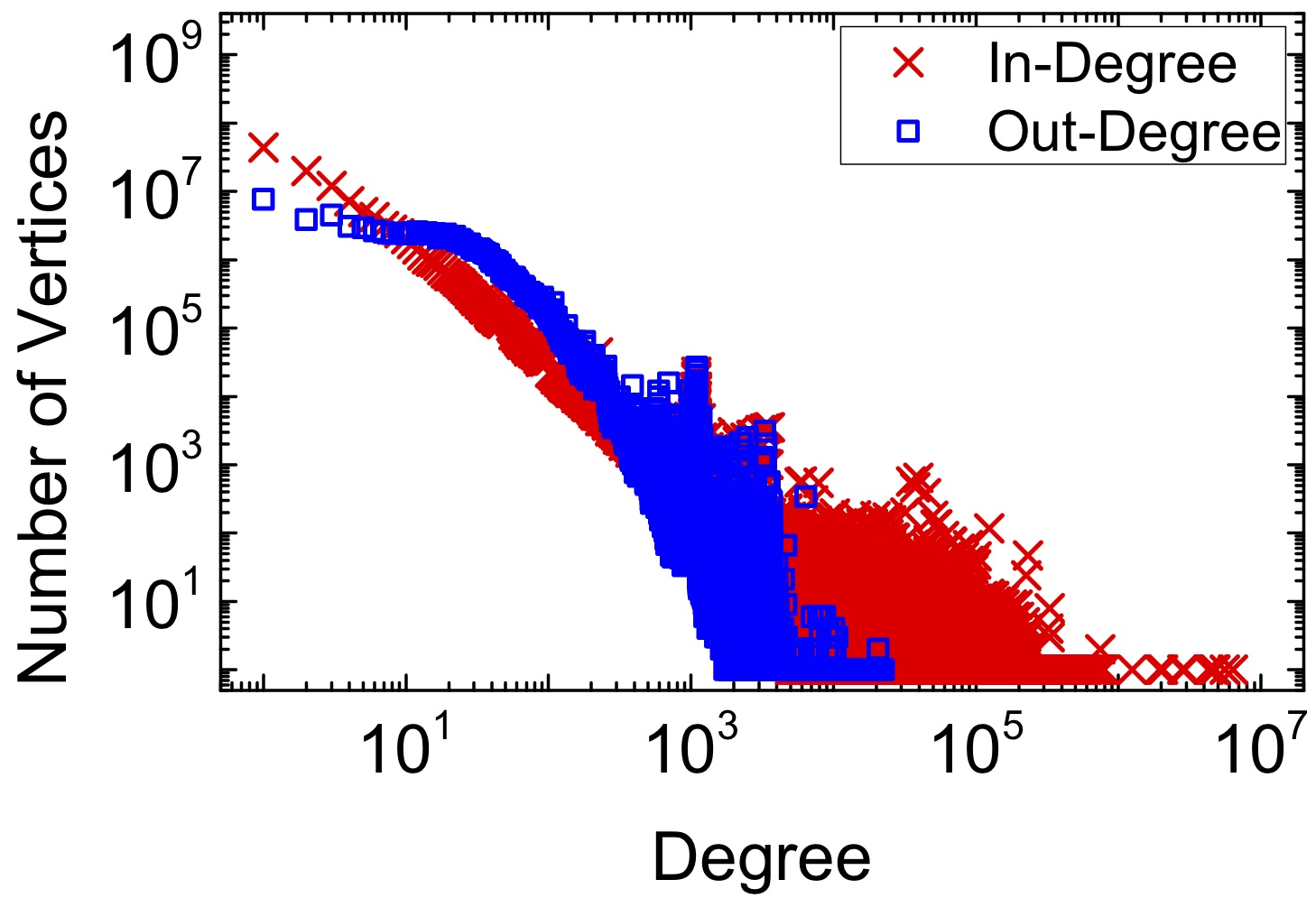}
    \subcaption{(b) UK-2007}
    \end{center}
    \end{minipage}
    \centering
    \begin{minipage}[t]{\minipagewidth}
    \begin{center}
    \includegraphics[width=\figurewidthFour]{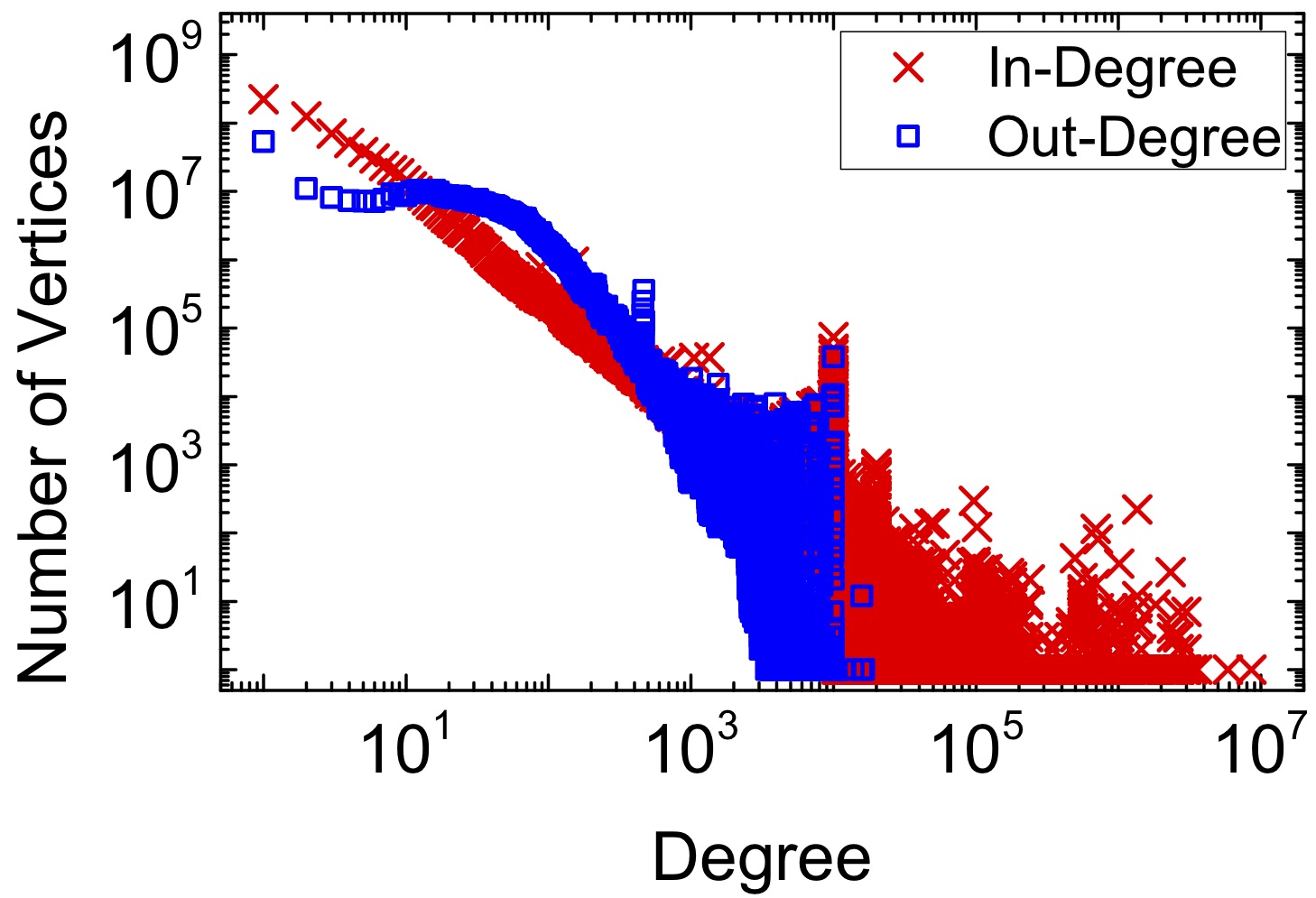}
    \subcaption{(c) UK-2014}
    \end{center}
    \end{minipage}
    \centering
    \begin{minipage}[t]{\minipagewidth}
    \begin{center}
    \includegraphics[width=\figurewidthFour]{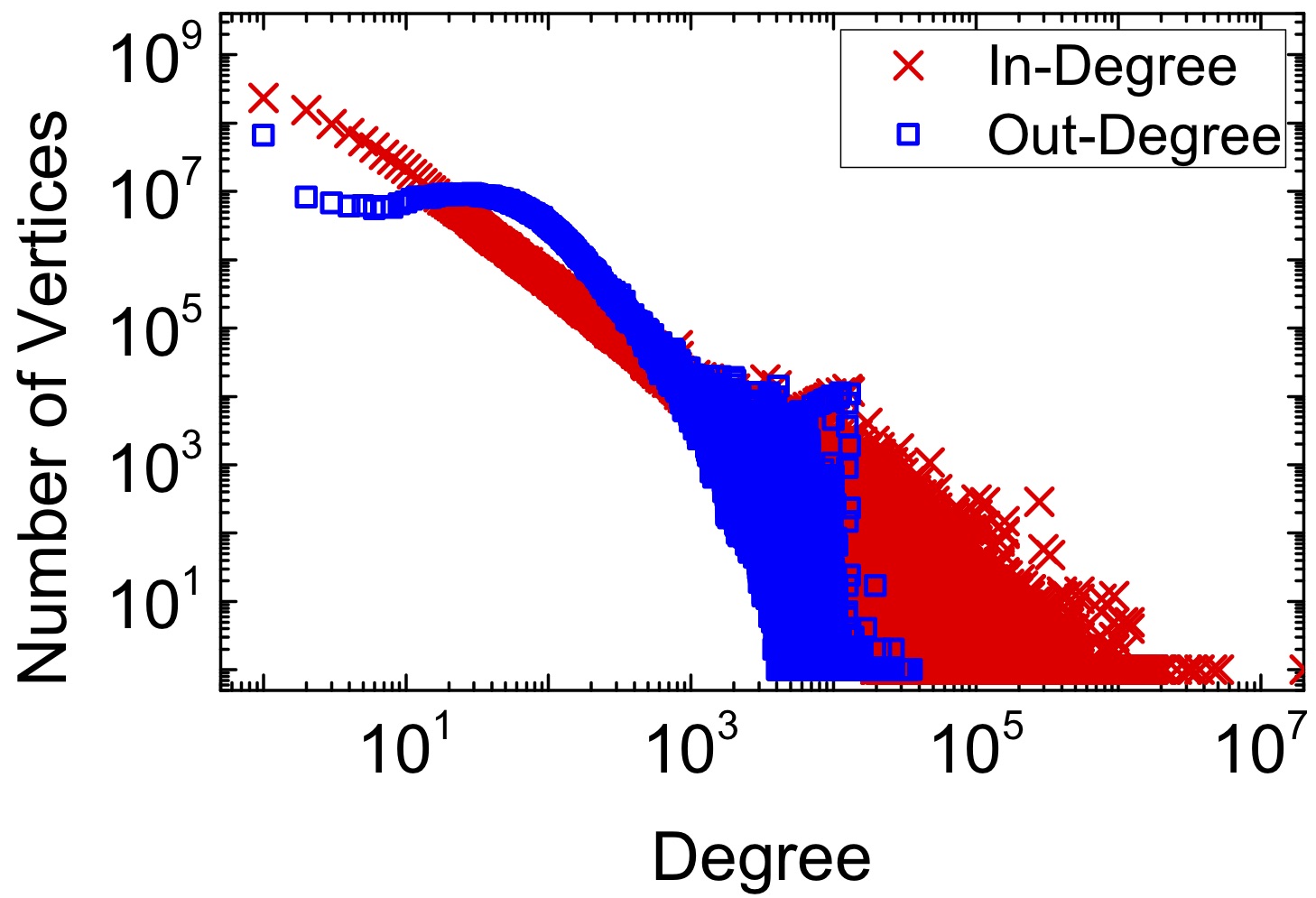}
    \subcaption{(d) EU-2015}
    \end{center}
    \end{minipage}
    \centering
    \caption{The in-degree and out-degree distribution of used graph datasets. All four graphs are power-law graphs: most vertices have relatively few neighbors while a few have many neighbors.}
\label{Fig: Input_Data}
\end{figure}

\renewcommand\arraystretch{1}
\begin{table}
\centering
\resizebox{0.49\textwidth}{!}{
\begin{threeparttable}{}
\caption{Graph datasets used in the experiments\tnote{1}.}
\begin{tabular}{@{}l c c c c c c@{}}
\thickhline
\textbf{Dataset} & \begin{tabular}[c]{@{}c@{}} \textbf{Vertex} \\ \textbf{Num} \end{tabular}   & \begin{tabular}[c]{@{}c@{}} \textbf{Edge} \\ \textbf{Num} \end{tabular}   & \begin{tabular}[c]{@{}c@{}} \textbf{Avg} \\ \textbf{Deg} \end{tabular}  & \begin{tabular}[c]{@{}c@{}} \textbf{Max} \\ \textbf{Indeg} \end{tabular} & \begin{tabular}[c]{@{}c@{}} \textbf{Max} \\ \textbf{Outdeg} \end{tabular} &  \begin{tabular}[c]{@{}c@{}} \textbf{Size} \\ \textbf{(CSV)} \end{tabular}  \\ \hline
\textbf{Twitter} & 42M   & 1.5B  & 35.3 &0.7M &770K & 25GB      \\
\textbf{UK-2007} & 134M   & 5.5B  & 41.2 &6.3M &22.4K & 93GB      \\   
\textbf{UK-2014} & 788M   & 47.6B & 60.4 &8.6M &16.3K  &0.9TB     \\
\textbf{EU-2015} & 1.1B & 91.8B & 85.7 &20M &35.3K & 1.7TB \\ \thickhline
\end{tabular}
\begin{tablenotes}
\item[1] All datasets is public on http://law.di.unimi.it/datasets.php.
\end{tablenotes}
\label{Tab: Datasets}
\end{threeparttable}
}
\end{table} 

In this section, we first evaluate the effect of GraphMP's selective scheduling and compressed edge caching.  Then, we compare the performance of GraphMP with GraphMat, which is a single-machine in-memory graph system. Next, we compare the performance of GraphMP with three popular single-machine out-of-core engines: GraphChi, X-Stream and GridGraph. Finally, we compare the performance of GraphMP with three distributed in-memory graph engines (Pregel+, PowerGraph and PowerLyra) and two distributed out-of-core systems ( GraphD and Chaos). We set up aforementioned distributed engines on 9  R720 servers connected by 10Gbps network. Each server has the same configuration with the server used to run single-machine graph engines.

\setlength{\minipagewidth}{0.235\textwidth}
\setlength{\figurewidthFour}{\minipagewidth}
\begin{figure} 
    \centering
    \begin{minipage}[t]{\minipagewidth}
    \begin{center}
    \includegraphics[width=\figurewidthFour]{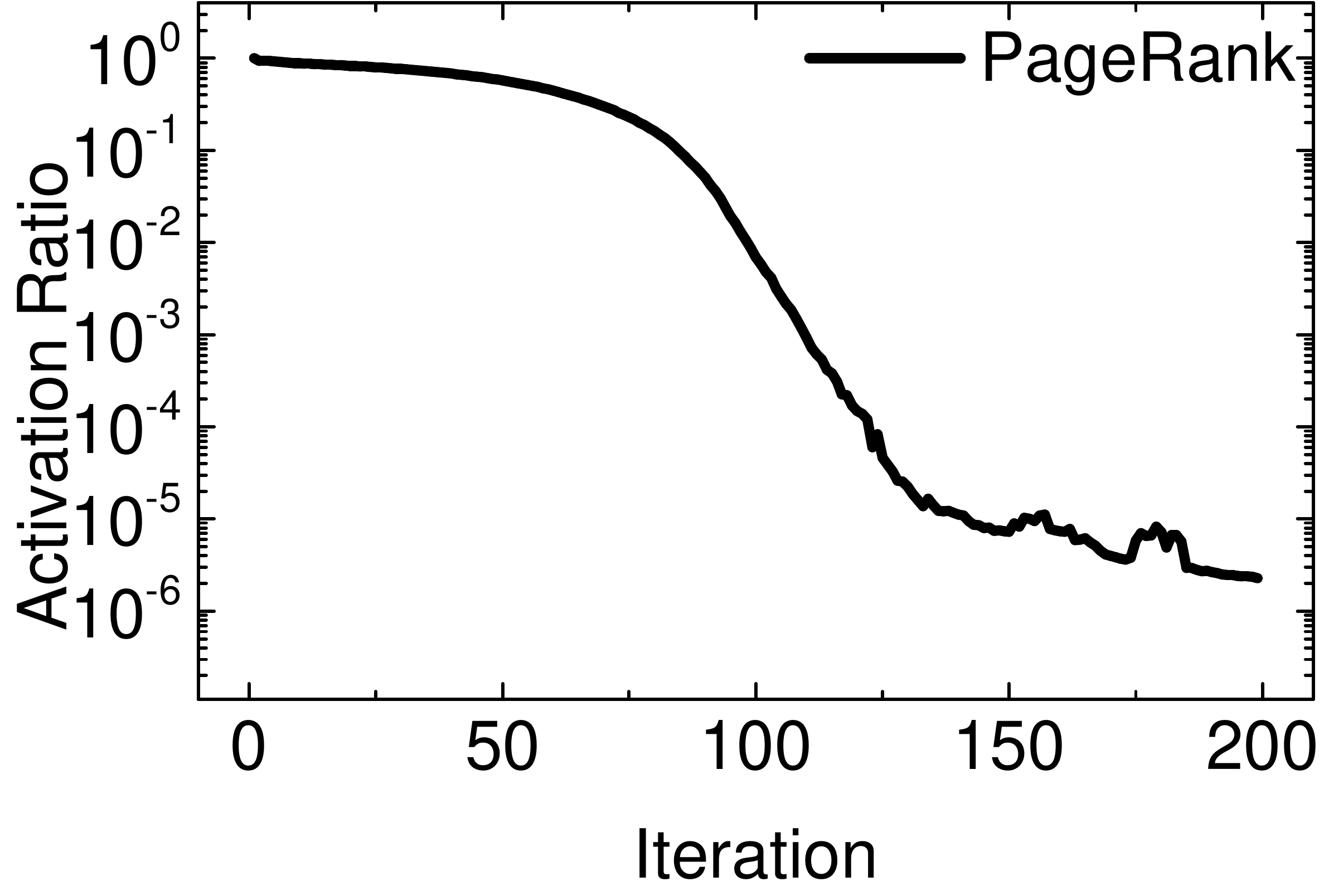}
    \subcaption{(a1) PageRank Vertex Activation Ratio}
    \end{center}
    \end{minipage}
    \centering
    \begin{minipage}[t]{\minipagewidth}
    \begin{center}
    \includegraphics[width=\figurewidthFour]{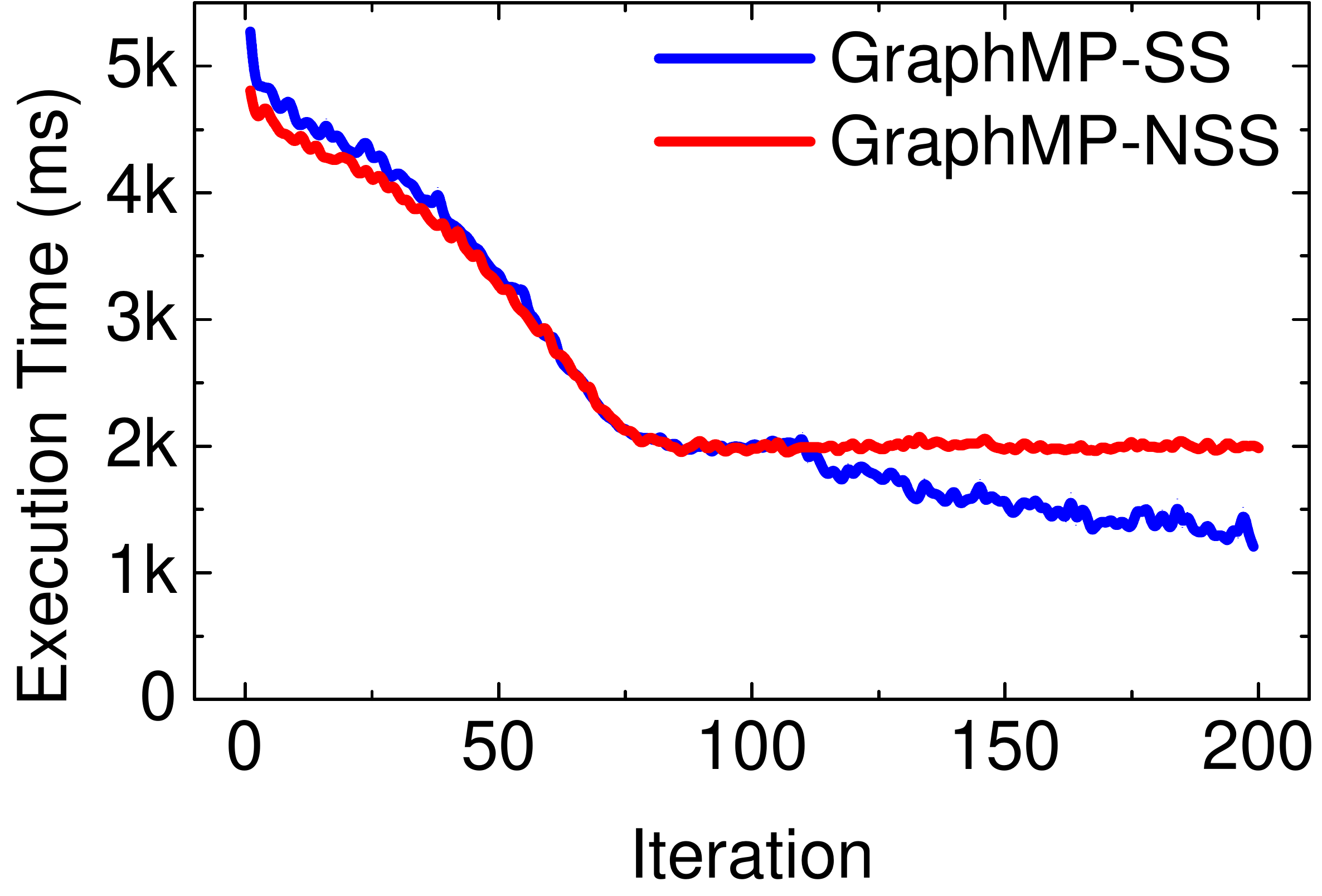}
    \subcaption{(a2) PageRank Execution Time}
    \end{center}
    \end{minipage}
    \centering
    \begin{minipage}[t]{\minipagewidth}
    \begin{center}
    \includegraphics[width=\figurewidthFour]{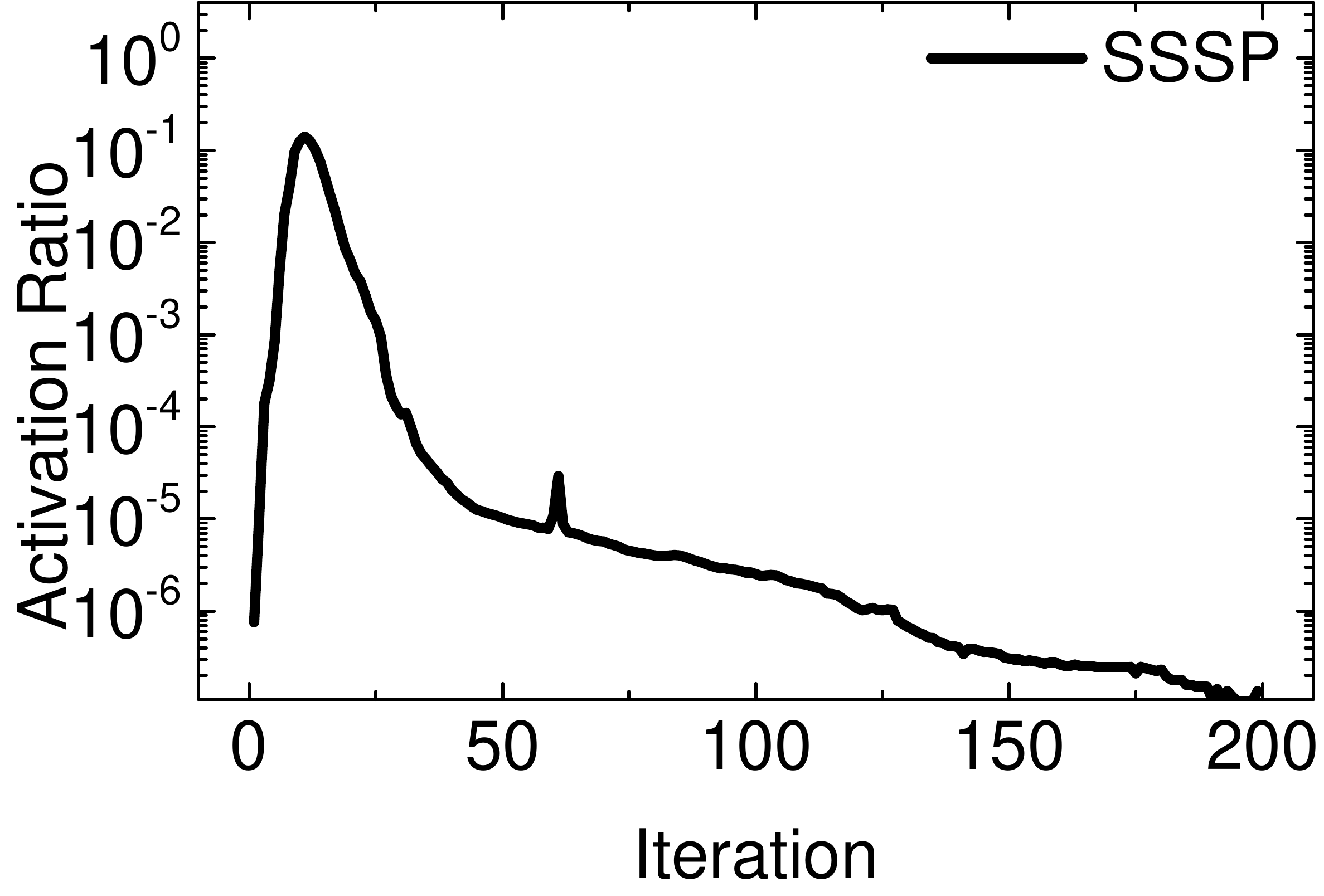}
    \subcaption{(b1) SSSP Vertex Activation Ratio}
    \end{center}
    \end{minipage}
    \centering
    \begin{minipage}[t]{\minipagewidth}
    \begin{center}
    \includegraphics[width=\figurewidthFour]{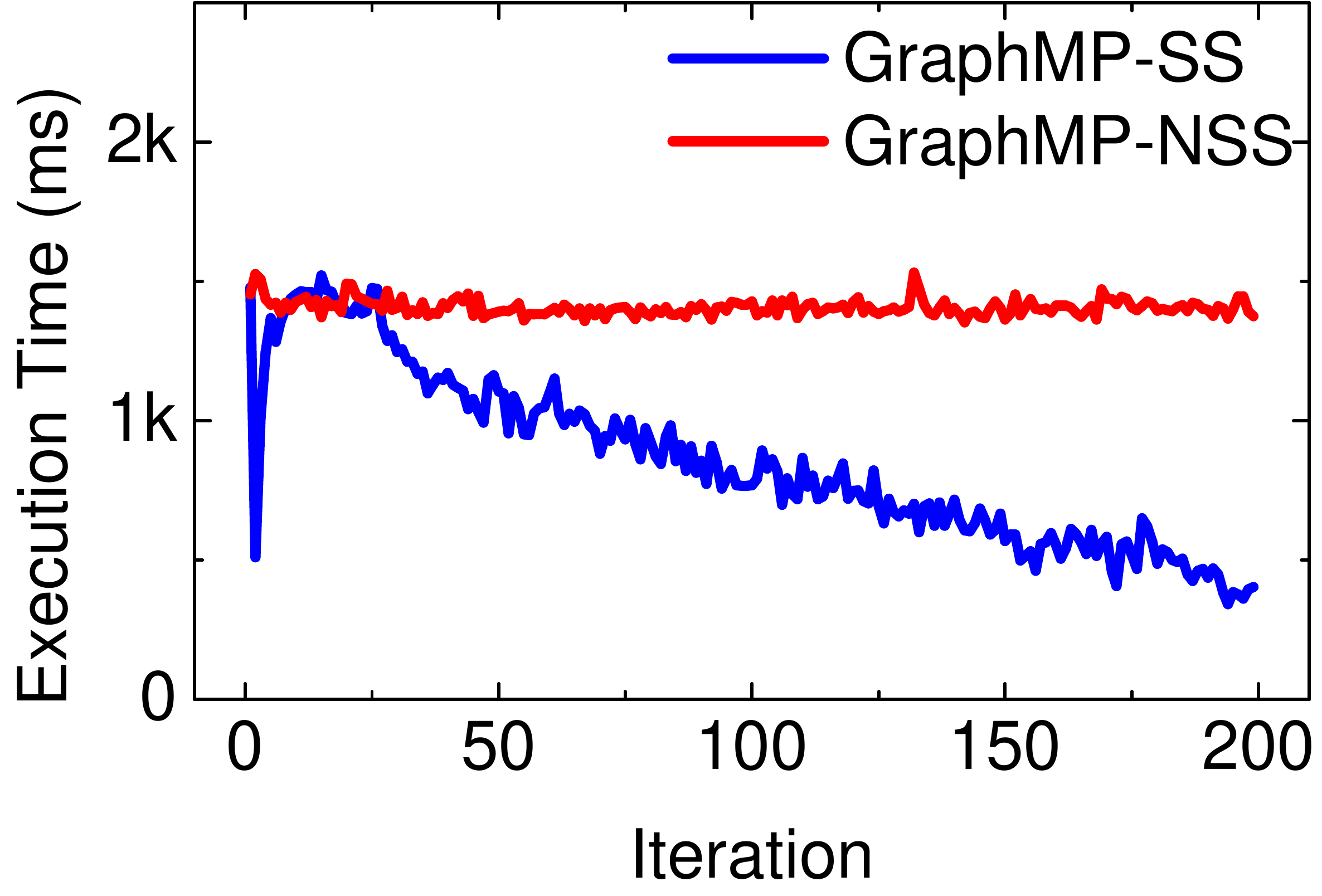}
    \subcaption{(b2) SSSP Execution Time}
    \end{center}
    \end{minipage}
    \centering
    \begin{minipage}[t]{\minipagewidth}
    \begin{center}
    \includegraphics[width=\figurewidthFour]{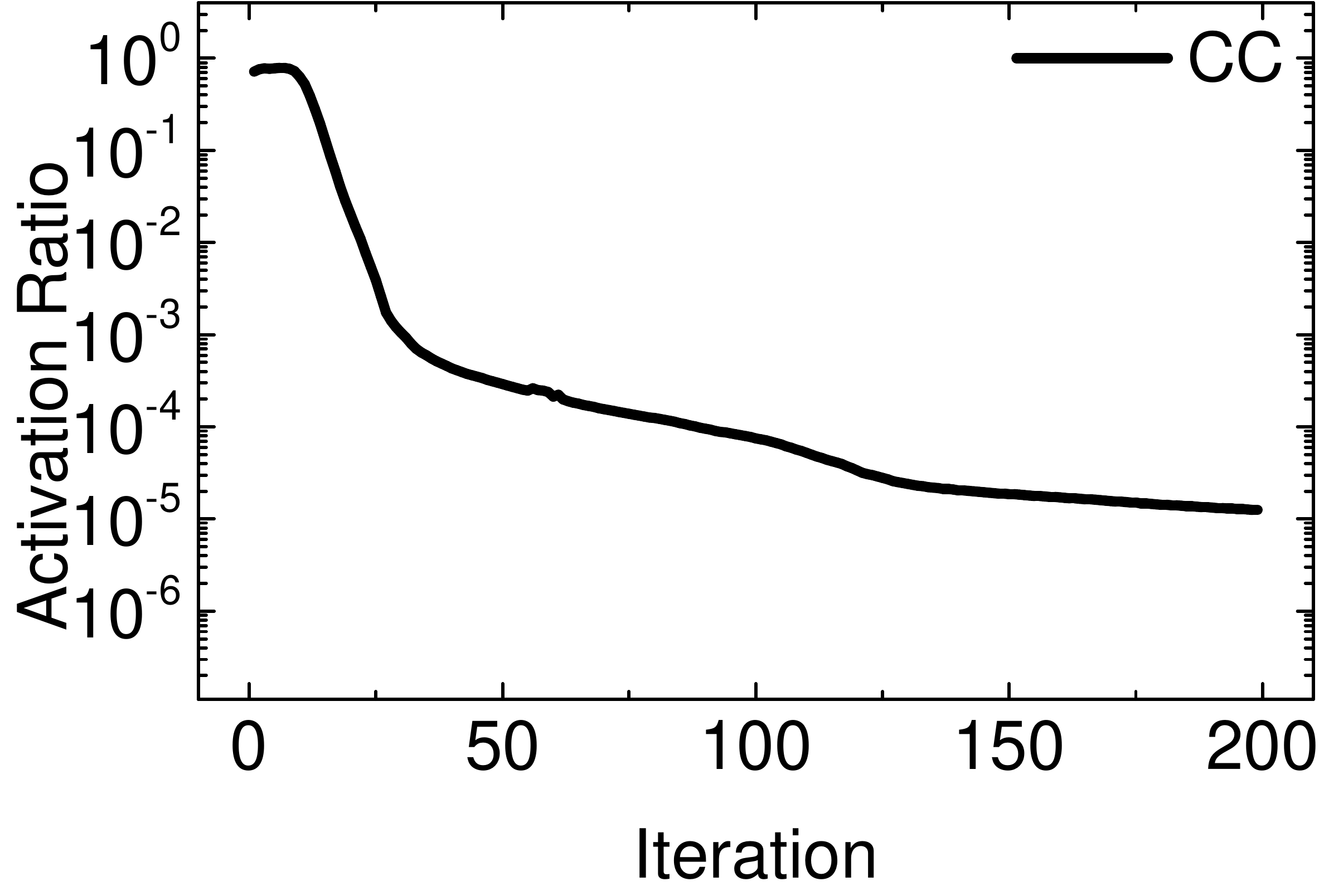}
    \subcaption{(c1) CC Vertex Activation Ratio}
    \end{center}
    \end{minipage}
    \centering
    \begin{minipage}[t]{\minipagewidth}
    \begin{center}
    \includegraphics[width=\figurewidthFour]{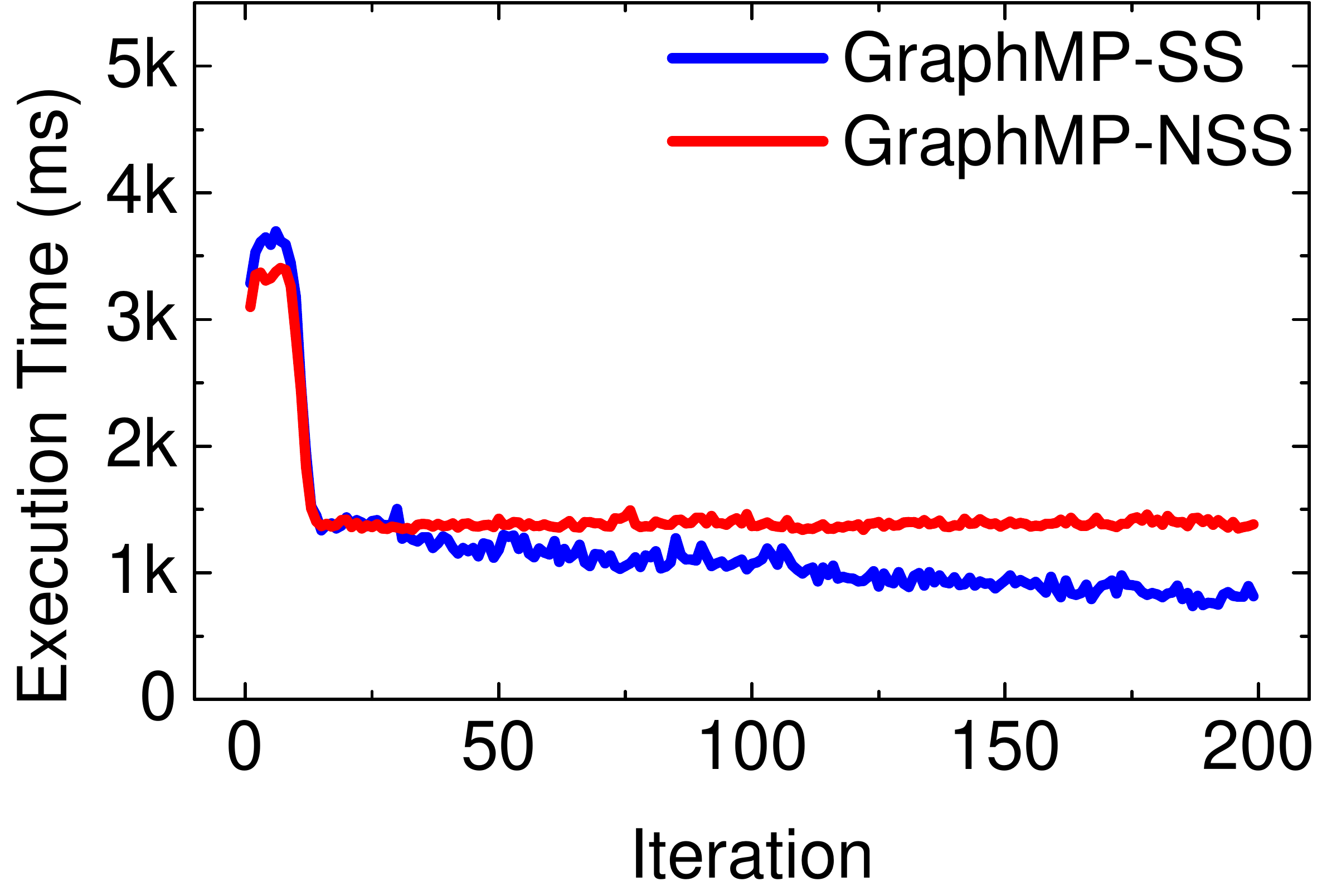}
    \subcaption{(c2) CC Execution Time}
    \end{center}
    \end{minipage}
    \caption{Effect of GraphMP's selective scheduling mechanism. GraphMP-SS enables selective scheduling. GraphMP-NSS disables the selective scheduling mechanism. We use UK-2007 as the input, and run PageRank, SSSP and CC on a single machine. The vertex activation ratio denotes the number of active vertices of an iteration.}
\label{Fig: Result_Selective}
\end{figure}

\subsection{Effect of Selective Scheduling}

In this set of experiments, we enable selective scheduling in  GraphMP-SS, so that it can use Bloom filters to detect and skip inactive shards. In GraphMP-NSS, we disable selective scheduling, so that it should process all shards in each iteration. To see the effect of GraphMP's selective scheduling, we run PageRank, SSP and CC on UK-2007 using GraphMP-SS and GraphMP-NSS, and compare their performance.   Figure \ref{Fig: Result_Selective} shows that GraphMP's selective scheduling  could improve the processing performance for all three applications.

As shown in Figure \ref{Fig: Result_Selective} (a1), when running PageRank on UK-2007, many vertices  converge quickly. After the 110-th iteration, less than $0.1\%$ of vertices update their values in an iteration. After that iteration, GraphMP-SS enables selective scheduling, and it continually reduces the execution time of an iteration. In particular, GraphMP-SS only uses $1.2$ seconds to execute the 200-th iteration. As a comparison, GraphMP-NSS roughly uses $2$ seconds per iteration after the 110-th iteration. In this case, selective scheduling could improve the processing performance of a single iteration by a factor of up to 1.67, and improve the overall performance by $5.8\%$.

From Figure \ref{Fig: Result_Selective} (b1) and (b2), we find that SSSP benefits a lot from GraphMP's selective scheduling mechanism. In this experiment, GraphMP updates more than $0.1\%$ of vertices  in a few iterations. Therefore,  GraphMP-SS could continuously reduce the computation time from the 15-th iteration, and uses $0.4$ seconds in the 200-th iteration. As a comparison, GraphMP-NSS roughly uses $1.4$ seconds per iteration. In this case,  GraphMP's selective scheduling mechanism could speed up the computation of an iteration by a factor of up to 2.86, and improve the overall performance of SSSP by $50.1\%$.

GraphMP's selective scheduling mechanism is enabled after the 31-th iteration of CC, as shown in Figure \ref{Fig: Result_Selective} (c1) and (c2). GraphMP-SS begins to outperform GraphMP-NSS from that iteration.  GraphMP-SS uses 0.8 seconds in the 200-th iteration, and  GraphMP-SS uses 1.4 seconds. In this case, GraphMP's selective scheduling mechanism reduces the computation time of an iteration by a factor of up to 1.75, and improves the overall performance of CC by $9.5\%$.

\setlength{\minipagewidth}{0.235\textwidth}
\setlength{\figurewidthFour}{\minipagewidth}
\begin{figure} 
    \centering
    \begin{minipage}[t]{\minipagewidth}
    \begin{center}
    \includegraphics[width=\figurewidthFour]{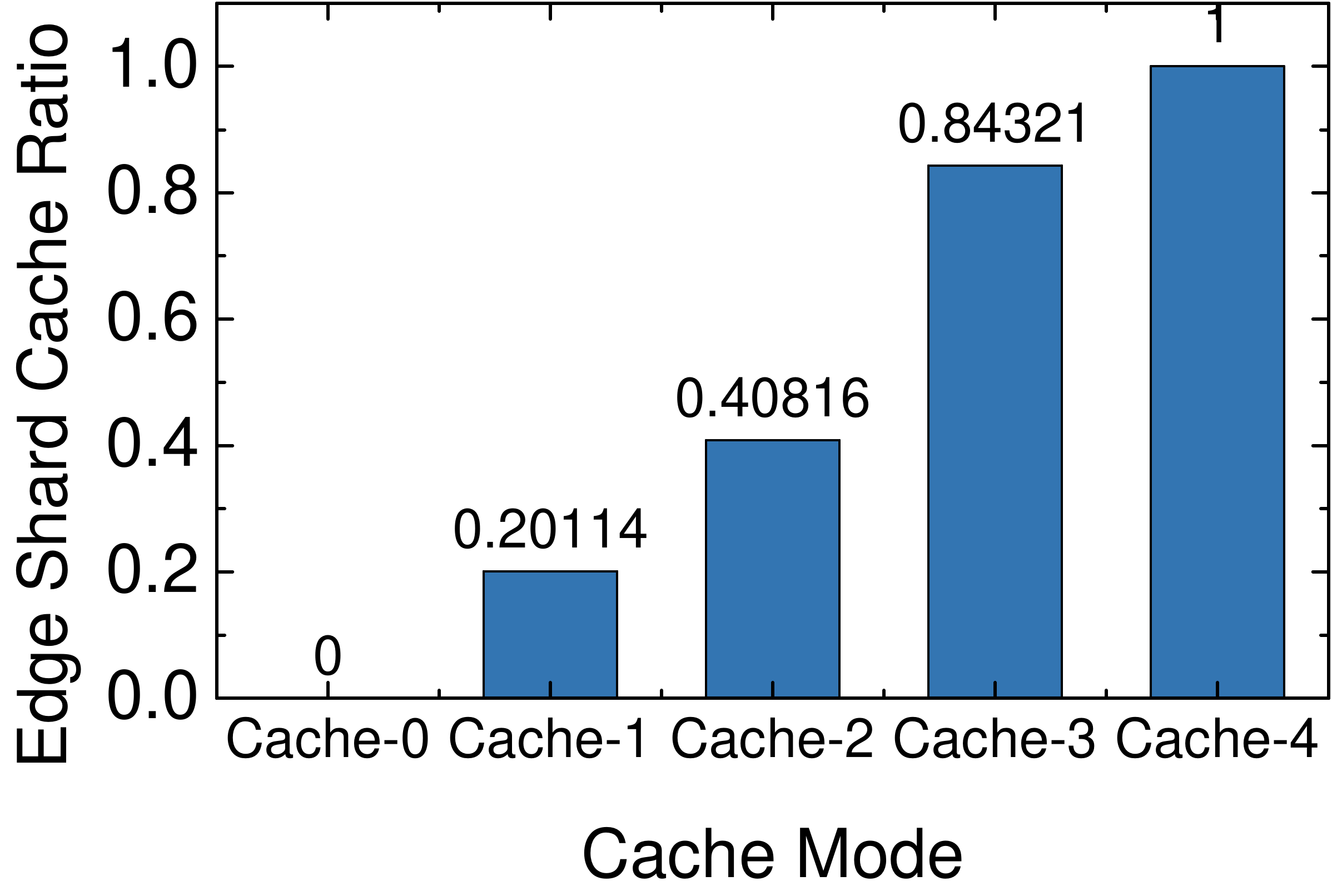} 
    \subcaption{(a) Edge Shard Cache Ratio}
    \end{center}
    \end{minipage}
    \centering
    \begin{minipage}[t]{\minipagewidth}
    \begin{center}
    \includegraphics[width=\figurewidthFour]{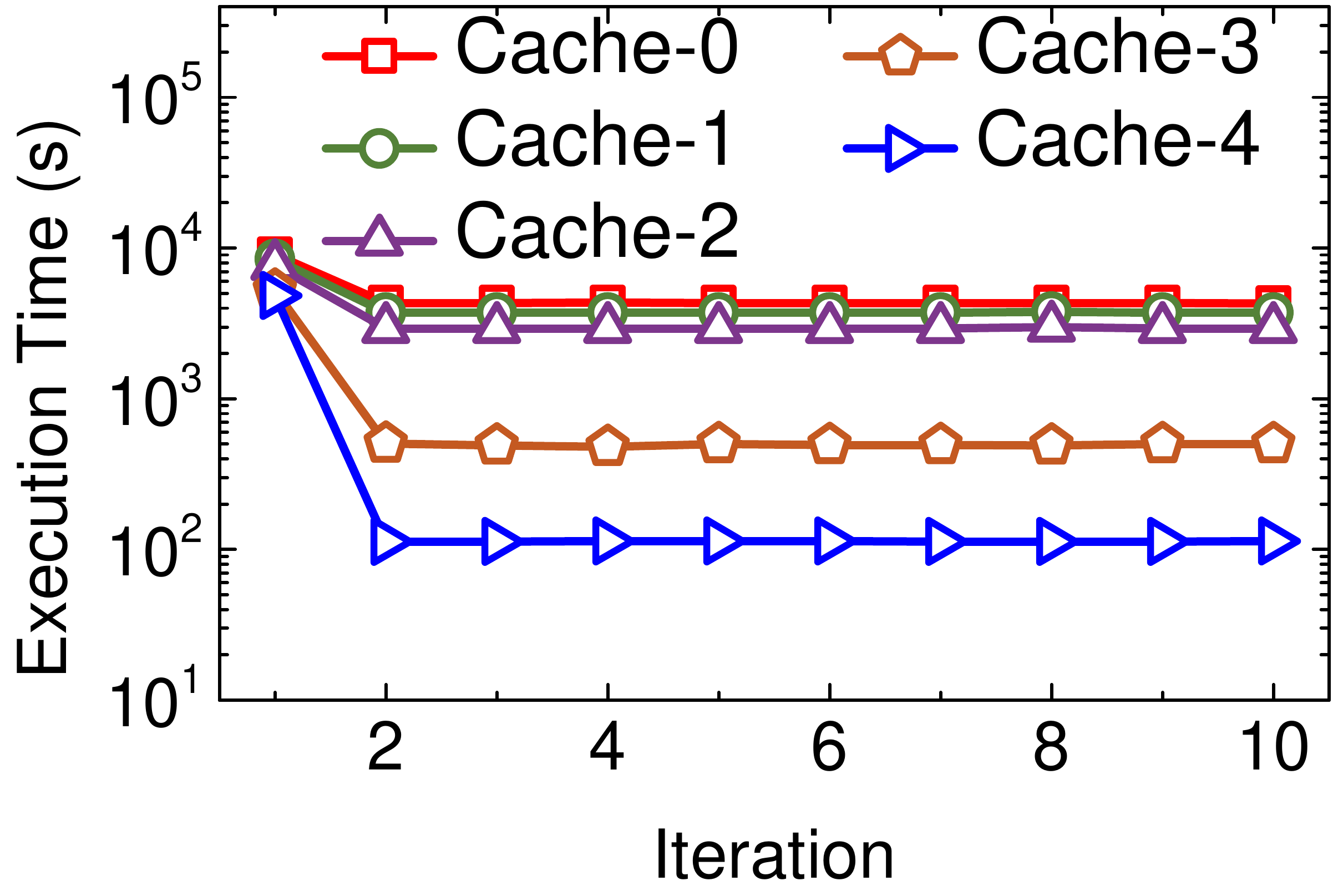}
    \subcaption{(b) PageRank Execution Time}
    \end{center}
    \end{minipage}
    \centering
    \begin{minipage}[t]{\minipagewidth}
    \begin{center}
    \includegraphics[width=\figurewidthFour]{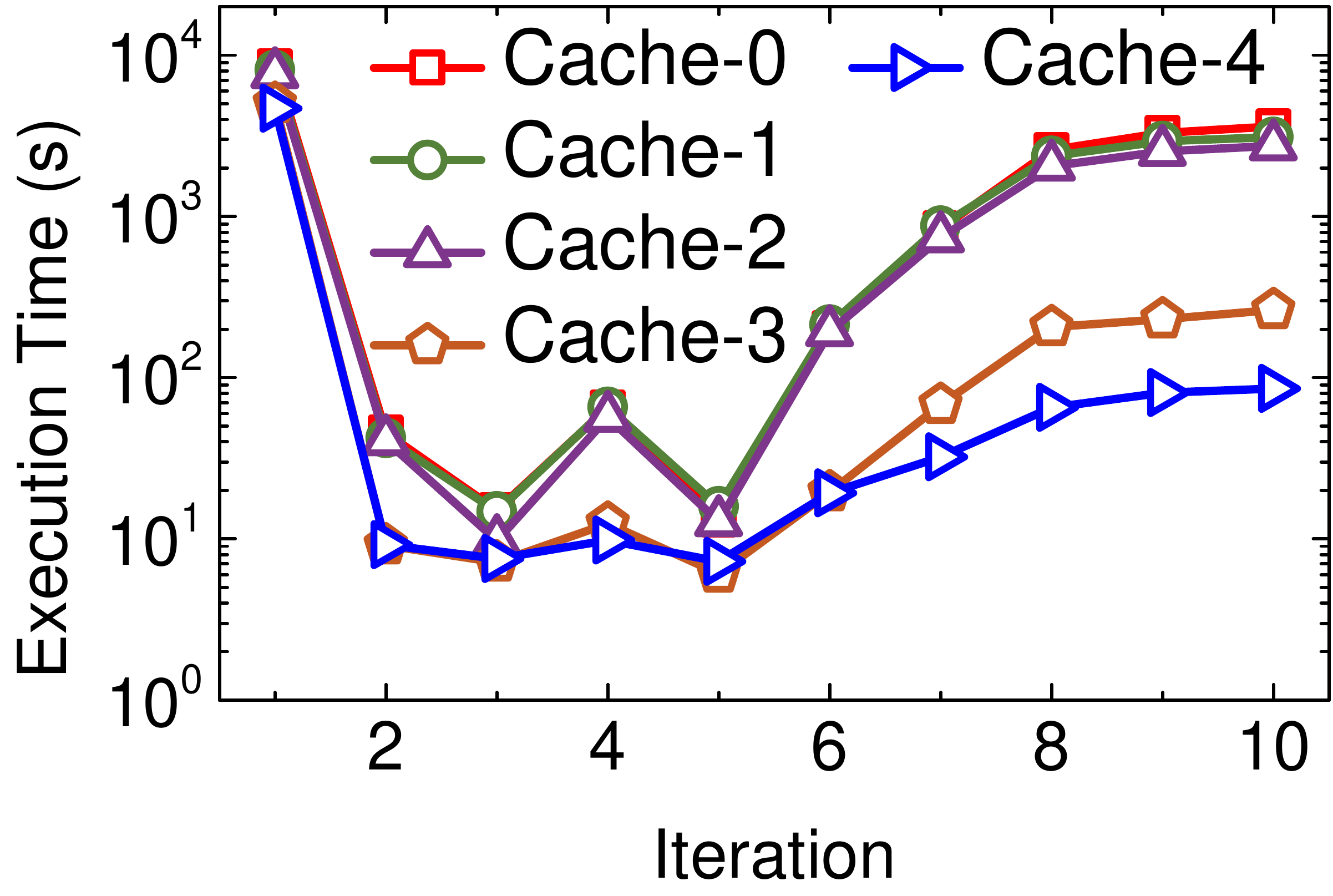}
    \subcaption{(c) SSSP Execution Time}
    \end{center}
    \end{minipage}
    \centering
    \begin{minipage}[t]{\minipagewidth}
    \begin{center}
    \includegraphics[width=\figurewidthFour]{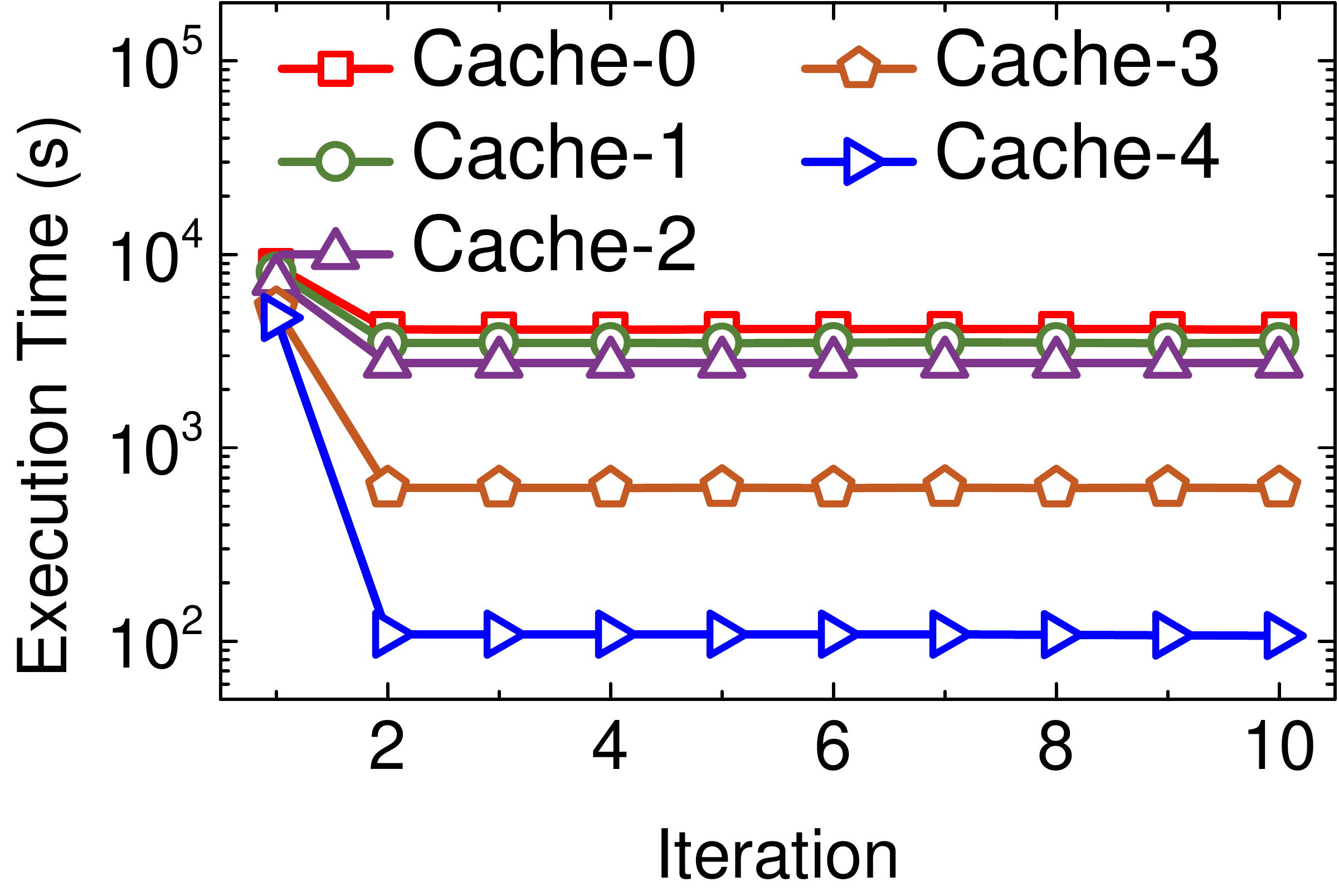}
    \subcaption{(d) CC Execution Time}
    \end{center}
    \end{minipage}
    \centering
    \caption{Effect of GraphMP's compressed edge caching.  We use EU-2015 as input, run PageRank, SSSP and CC on GraphMP with different cache modes, and capture the execution time of first 10 iterations.}
\label{Fig: Result_Cache}
\end{figure}

\subsection{Effect of Compressed Edge Caching}

To see the effect of GraphMP's compressed edge caching, we run PageRank, SSP and CC on EU-2015 using GraphMP with different cache modes, and compare their performance. 

As shown in Figure \ref{Fig: Result_Cache}(a), when using compressors with higher compression rate, GraphMP could cache more edge shards in memory. Specifically, GraphMP (Cache-0) could cache about 20\% of edge shards in memory without data compressing. In the same testbed, GraphMP (Cache-3) could cache about 84.3\% of edge shards and reduce disk I/O cost by using zlib-1 to compress cached edge shards. 
GraphMP (Cache-4) could cache all edge shards by compressing edge shards with zlib-3. In this case, there is even no disk I/O cost after loading all edge shards into memory.

From Figure \ref{Fig: Result_Cache}(b), (c) and (d), we can see that GraphMP's compressed edge caching  could significantly improve  graph processing performance. Since GraphMP should access all graph data from disk for filling edge cache and constructing Bloom filters during the first iteration, it takes more time to complete this iteration than others. Figure \ref{Fig: Result_Cache}(b) shows that GraphMP roughly takes $48650$ seconds to complete the first 10 iterations of PageRank with cache-0. The corresponding values of cache-1, cache-2, cache-3 and cache-4 are $42075$, $34077$, $9678$ and $5868$ seconds, respectively. In this case, GraphMP's compressed edge caching can speed up PageRank by $8.3$. When running SSSP, cache-1, cache-2, cache-3 and cache-4 could speed up  the application by $1.1$, $1.2$, $3.3$ and $3.8$ respectively, compared to cache-0, as shown in Figure \ref{Fig: Result_Cache}(c). In this experiment, GraphMP has less execution time from iteration 2 to iteration 6 due to selective scheduling: only a small portion of shards with active vertices are processed.
As shown in Figure \ref{Fig: Result_Cache}(d), CC also benefits from  compressed edge caching. More specifically, cache-1, cache-2, cache-3 and cache-4 could speed up CC by a factor of $1.2$, $1.4$, $4.2$ and $8.1$ respectively, when compared to cache-0.

\subsection{GraphMP vs. GraphMat}

We compare the performance of GraphMP with GraphMat, which is an in-memory graph processing system. GraphMat maps vertex-centric programs to sparse vector-matrix multiplication (SpMV) operations, and  leverages sparse linear algebra libraries and techniques to improve the performance of large-scale graph computation. The results show that GraphMP has competitive performance to GraphMat.

\setlength{\minipagewidth}{0.485\textwidth}
\setlength{\figurewidthFour}{\minipagewidth}
\begin{figure} 
    \centering
    \begin{minipage}[t]{\minipagewidth}
    \begin{center}
    \includegraphics[width=\figurewidthFour]{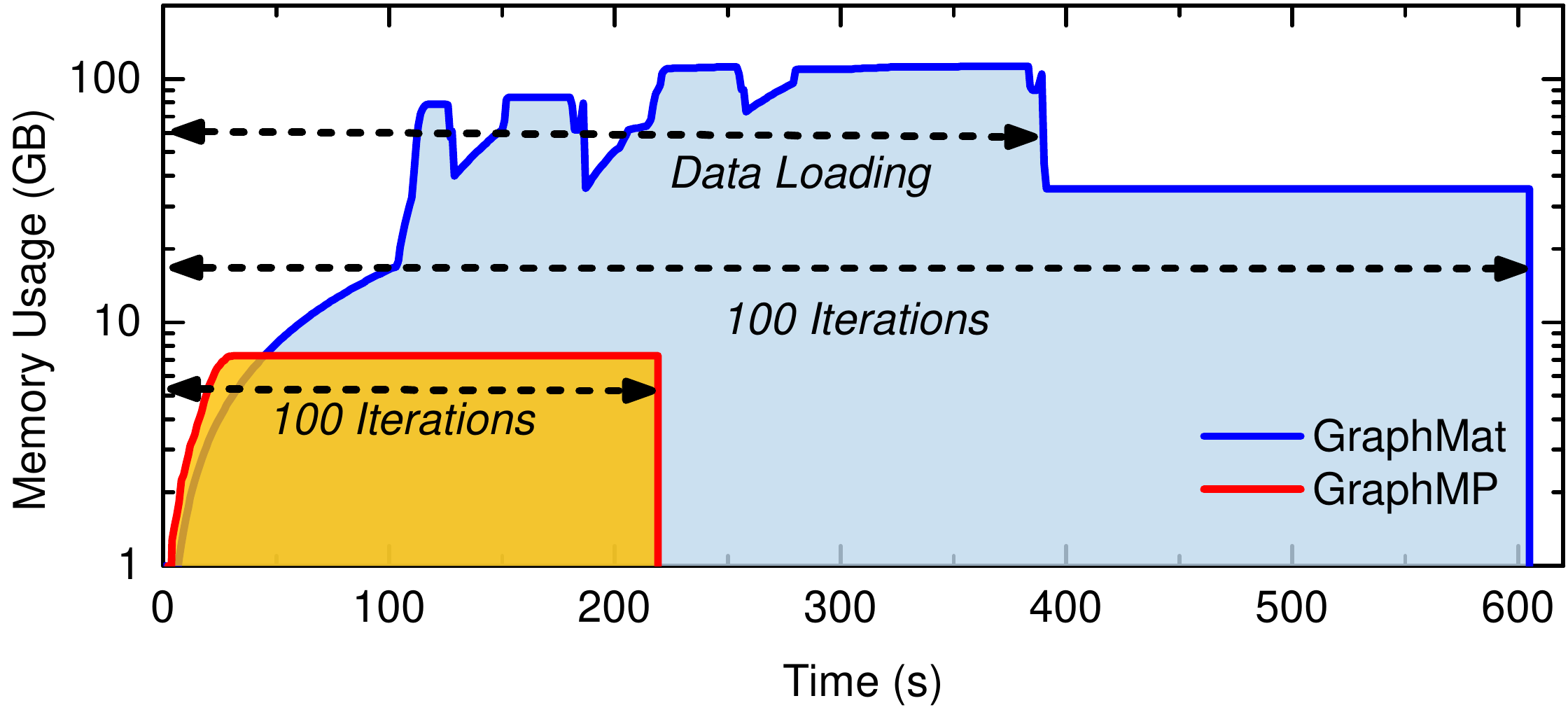}
    \end{center}
    \end{minipage}
    \centering
    \caption{Performance comparison between GraphMP and GraphMat. In this experiment, we run PageRank on the Twitter dataset.}
\label{Fig: Compare_Mat}
\end{figure}

GraphMat could not process UK-2007, UK-2014 and EU-2015 in our machine with 128GB memory.  At the beginning of a graph application, GraphMat  loads all vertices and edges into memory, and manages them with required data structures. As shown in Figure \ref{Fig: Compare_Mat}, when running PageRank on Twitter, GraphMat uses up to $122$GB memory for data loading. When processing UK-2007, UK-2014 and EU-2015 in our machine, GraphMat can easily crash  caused by out-of-memory. As a comparison, GraphMP could efficiently process all 4 datasets in a single machine.

From Figure \ref{Fig: Compare_Mat}, we can find that GraphMat's data loading phases use $390$ seconds to load the Twitter dataset into memory. Since GraphMat does not require the input graph's edges to be ordered, there is an expensive sorting process to build required data structures for SpMV  during the data loading phase. As a comparison,  GraphMP uses a separated data preprocessing stage to sort and group the input graph's edges, and could reuse these data in different applications. Thus, GraphMP uses $7.3$GB memory (including Bloom filters and edge cache) to run PageRank on Twitter, and takes about 30 seconds to complete the first iteration, which contains the  compressed edge cache filling time and  Bloom filter construction time. In addition, GraphMP needs additional $340.2$ seconds for data preprocessing.

\setlength{\minipagewidth}{0.235\textwidth}
\setlength{\figurewidthFour}{\minipagewidth}
\begin{figure} 
    \centering
    \begin{minipage}[t]{\minipagewidth}
    \begin{center}
    \includegraphics[width=\figurewidthFour]{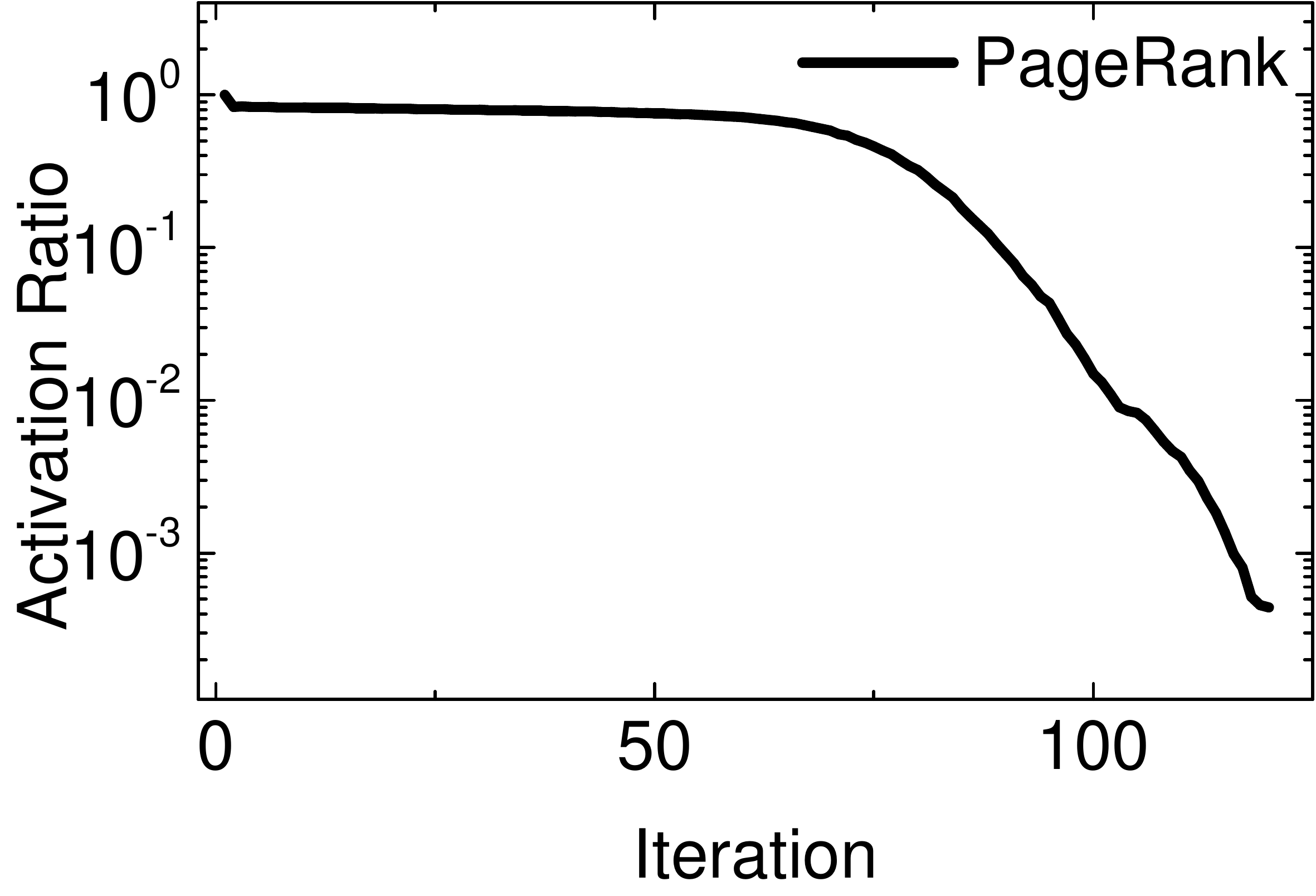}
    \subcaption{(a1) PageRank Vertex Activation Ratio}
    \end{center}
    \end{minipage}
    \centering
    \begin{minipage}[t]{\minipagewidth}
    \begin{center}
    \includegraphics[width=\figurewidthFour]{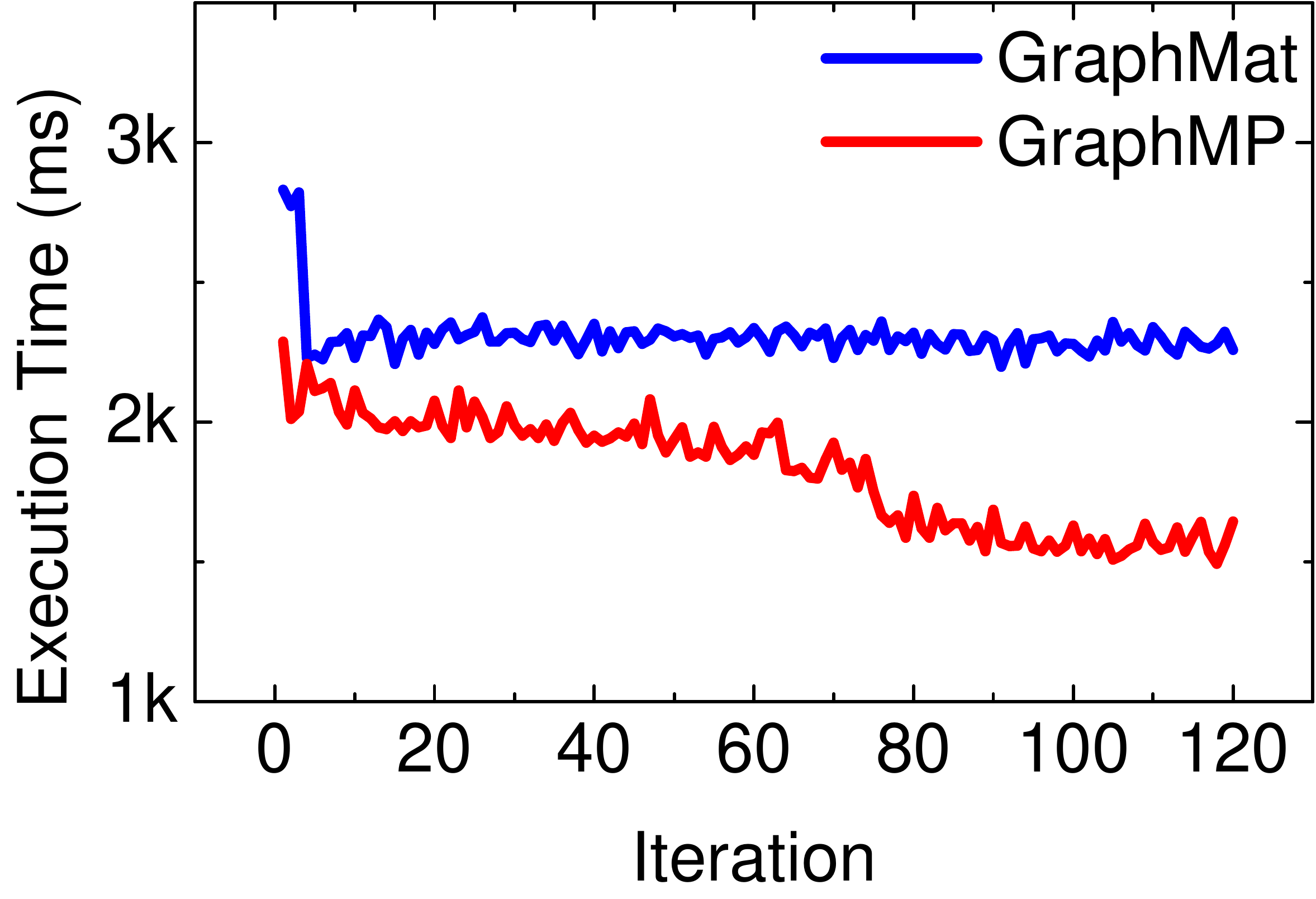}
    \subcaption{(a2) PageRank Execution Time}
    \end{center}
    \end{minipage}
    \centering
    \begin{minipage}[t]{\minipagewidth}
    \begin{center}
    \includegraphics[width=\figurewidthFour]{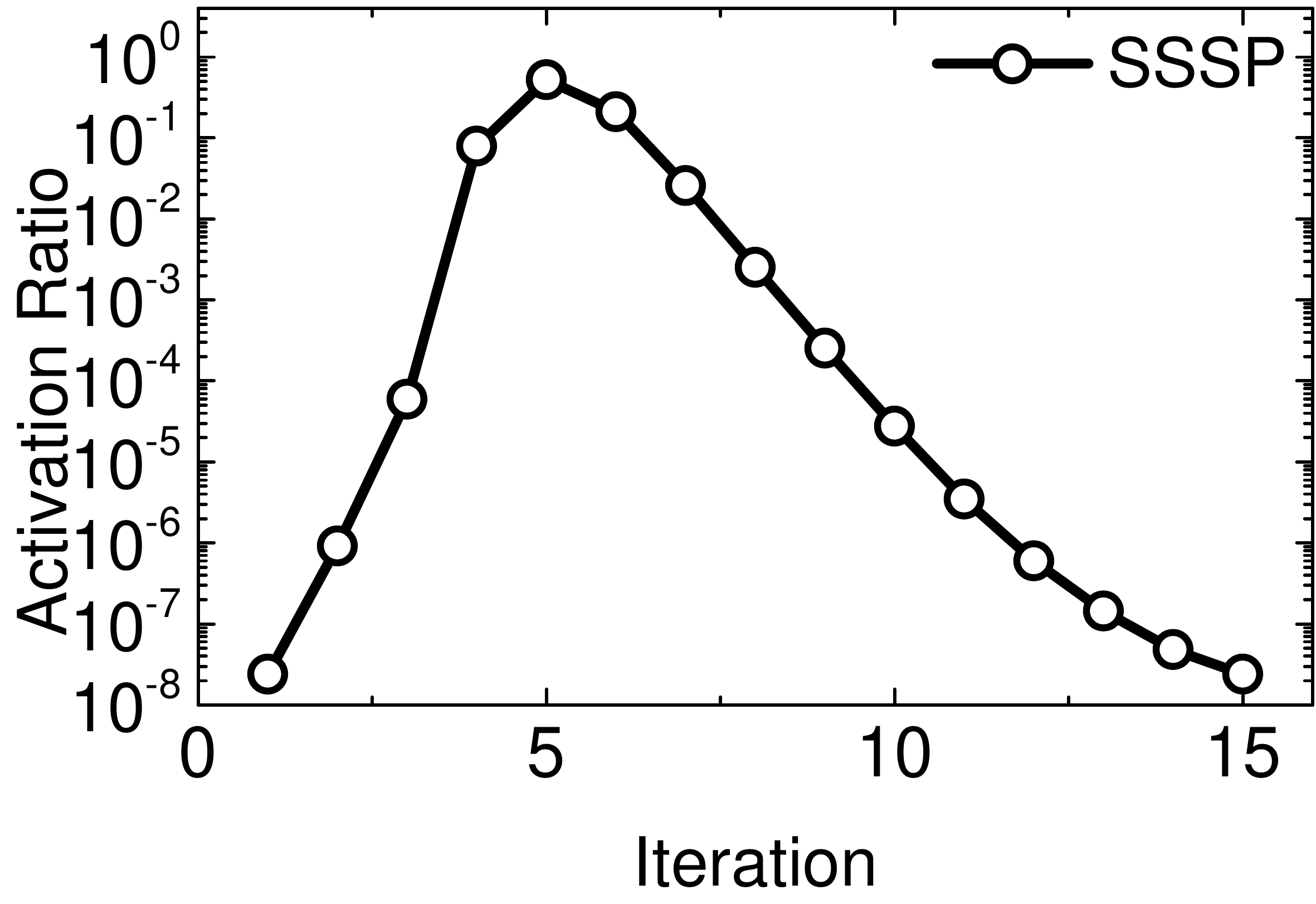}
    \subcaption{(b1) SSSP Vertex Activation Ratio}
    \end{center}
    \end{minipage}
    \centering
    \begin{minipage}[t]{\minipagewidth}
    \begin{center}
    \includegraphics[width=\figurewidthFour]{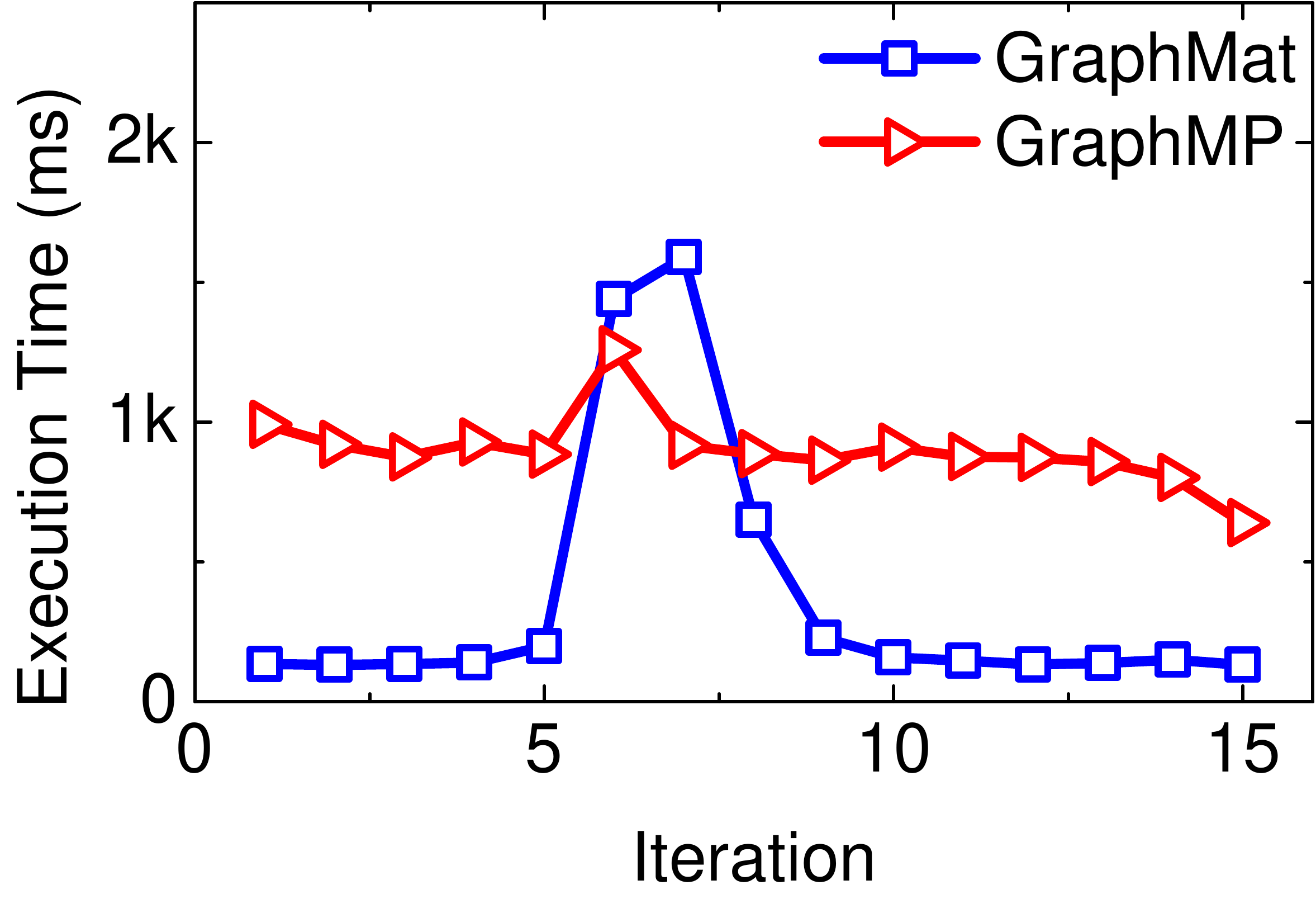}
    \subcaption{(b2) SSSP Execution Time}
    \end{center}
    \end{minipage}
    \centering
    \begin{minipage}[t]{\minipagewidth}
    \begin{center}
    \includegraphics[width=\figurewidthFour]{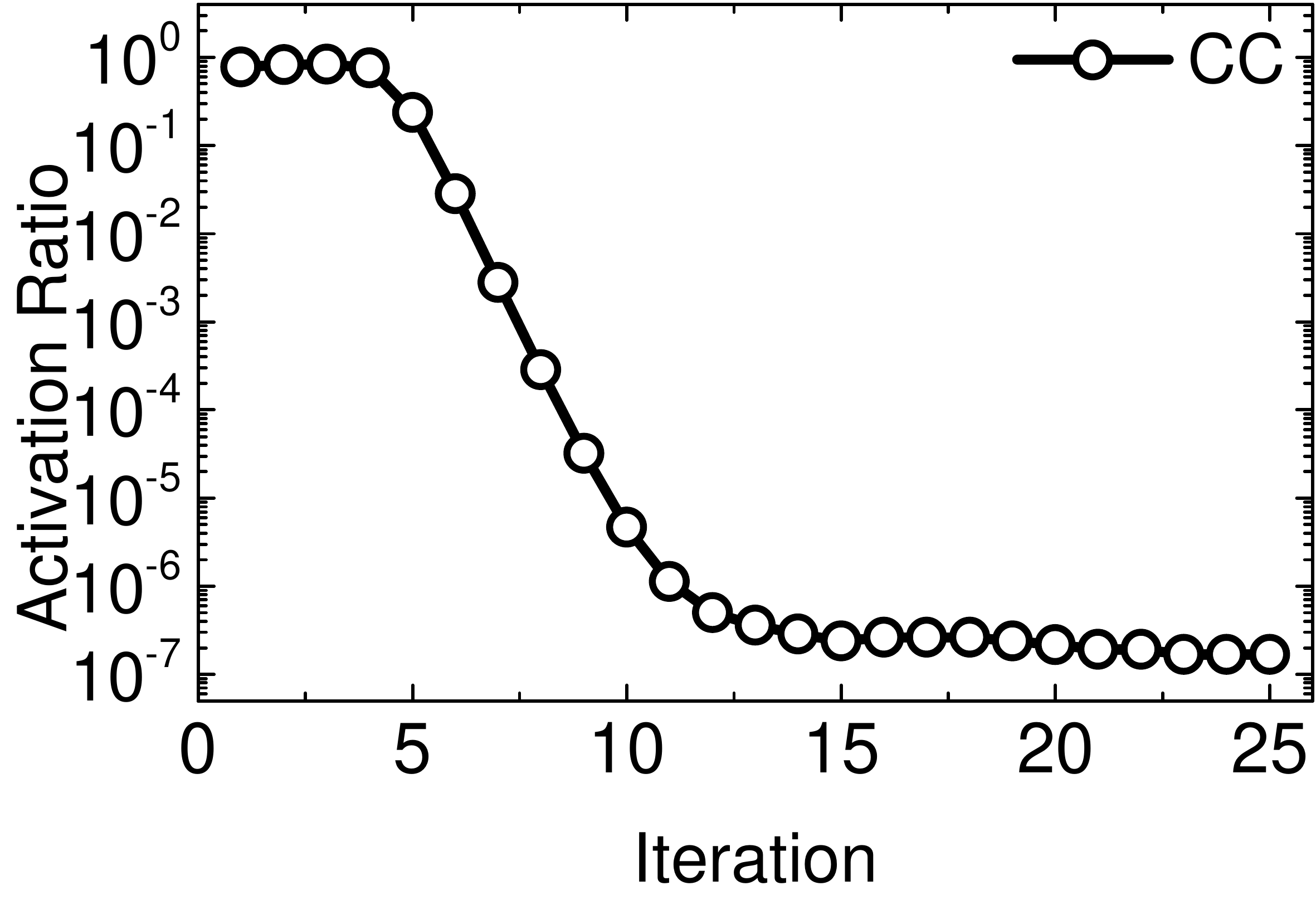}
    \subcaption{(c1) CC Vertex Activation Ratio}
    \end{center}
    \end{minipage}
    \centering
    \begin{minipage}[t]{\minipagewidth}
    \begin{center}
    \includegraphics[width=\figurewidthFour]{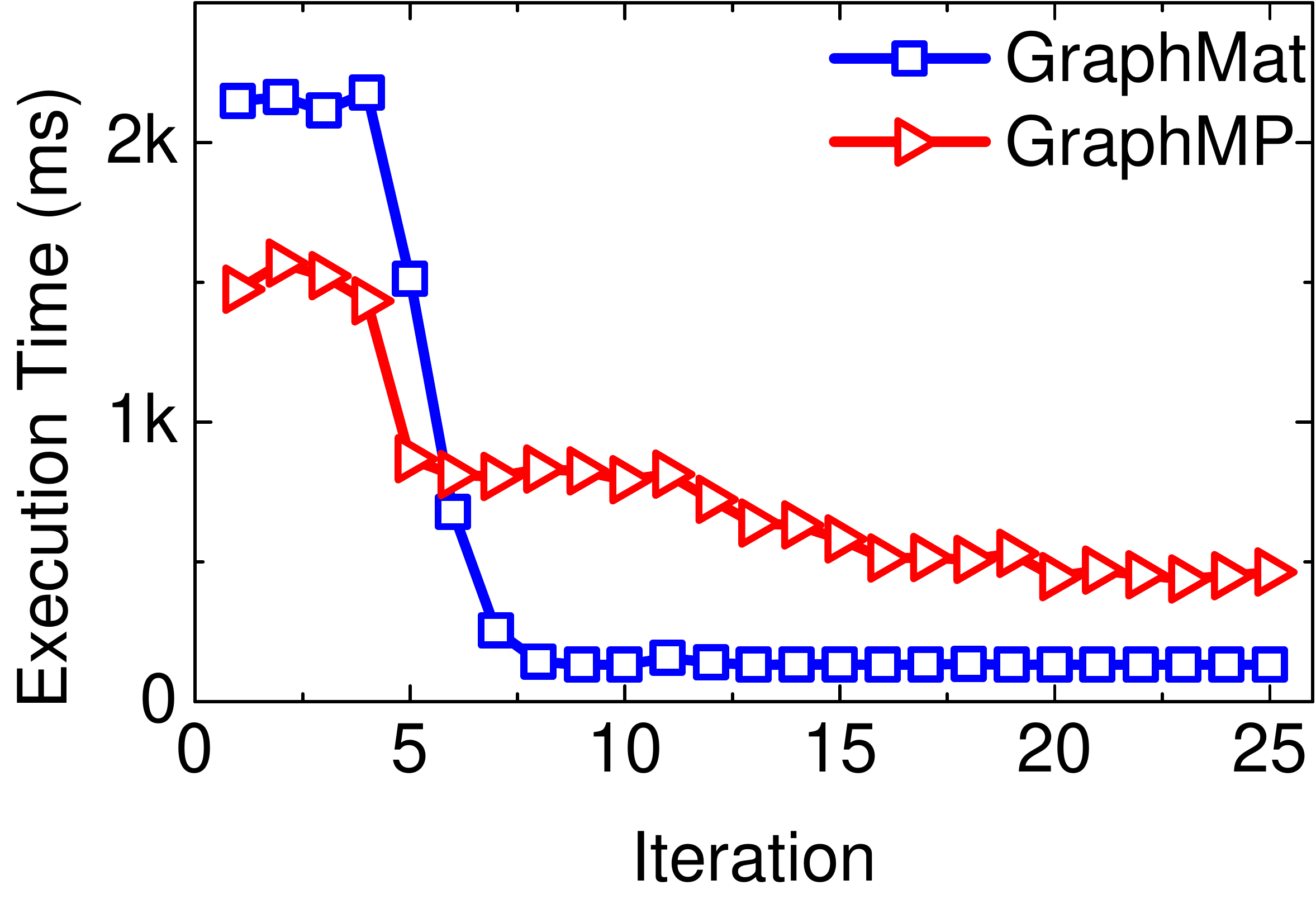}
    \subcaption{(c2) CC Execution Time}
    \end{center}
    \end{minipage}
    \caption{Performance comparison between GraphMP and GraphMat to run PageRank, SSSP and CC on the Twitter dataset. Vertex activation ratio denotes the ratio of active vertices of an iteration. In this figure, the first iteration's execution time does not include data loading time or initialization time for fair comparison.}
\label{Fig: Compare_Mat2}
\end{figure}

Since GraphMP performs costly edge sorting in a separated data preprocessing stage and GraphMat sorts edges at the beginning of an application, we compare the performance of GraphMat with GraphMP in two cases. In the first case, we do not consider data loading overhead for both systems. In GraphMat, we start to measure the execution time when all vertices and edges are loaded into memory and all data structures are constructed. In GraphMP, we do not consider the time used for filling edge cache and constructing Bloom filters. Figure \ref{Fig: Compare_Mat2} shows  vertex activation ratio  and  running time  of each iteration when running PageRank, SSSP and CC on Twitter.
We can see that GraphMat takes $28$ seconds to complete the first 120 iterations of PageRank, and GraphMP uses $22$ seconds. GraphMat takes $1.3$ seconds to complete the first 15 iterations of SSSP, while GraphMP needs $9.9$s. When running CC, GraphMat takes $1.5$ seconds to complete the first $25$ iterations, while GraphMP uses $2.1$ seconds. 
In the second case, we consider data loading and data preprocessing overhead in both systems. Specifically, we add  data preprocessing time of GraphMP to the total execution time. In this case,  GraphMat takes $418$ seconds for running PageRank, while GraphMP uses $366$ seconds. GraphMat takes $349$ seconds to run SSSP, while GraphMP needs $361$s. When running CC in our testbed, GraphMat takes $382$ seconds, while GraphMP uses $373$ seconds.

\begin{table*}[]
\centering
\caption{Performance comparison between GraphMP with other systems (application: PageRank; time unit: minutes; time collection: first 10 iterations).}
\label{Tab: perf_all_pagerank}
\begin{tabular}{@{}c|ccc|ccc|cc|cc@{}}
\thickhline
\multirow{2}{*}{\textbf{Dataset}} & \multicolumn{3}{c|}{\textbf{Single-Machine Out-of-Core}} & \multicolumn{3}{c|}{\textbf{Distributed In-Memory}} & \multicolumn{2}{@{}c@{}|}{\textbf{\;\;Distributed Out-of-Core\;\;}} & \multicolumn{2}{c}{\textbf{GraphMP}} \\ 
\multicolumn{1}{l|}{} & \textbf{GraphChi} & \textbf{X-Stream} & \textbf{GridGraph} & \textbf{Pregel+} & \textbf{PowerGraph} & \textbf{PowerLyra} & \textbf{\;\,GraphD\;\,} & \textbf{\;\,Chaos\;\,} & \textbf{NoCache} & \textbf{Cache} \\ \hline
\textbf{Twitter} & 7.35  & 22.62 & 2.73 & 1.15 & 0.92 & 0.78 & 2.02 & 3.64 & 0.76 & 0.67 \\
\textbf{UK-2007} & 18.71 & 148.27 & 7.57 & 5.49 & 3.22 & 2.55 & 13.38 & 14.11 & 2.99 & 2.90 \\
\textbf{UK-2014} & 473.33 & 1413.35 & 675.73 & - & - & - & 543.82 & 389.90 & 336.45 & 46.36 \\
\textbf{EU-2015} & 970.67 & 2856.78 & 1162.63 & - & - & - & 2267.43 & 751.09 & 797.98 & 94.48 \\ \thickhline
\end{tabular}
\end{table*}

\begin{table*}[]
\centering
\caption{Performance comparison between GraphMP with other systems (application: SSSP; time unit: minutes; time collection: first 10 iterations).}
\label{Tab: perf_all_sssp}
\begin{tabular}{@{}c|ccc|ccc|cc|cc@{}}
\thickhline
\multirow{2}{*}{\textbf{Dataset}} & \multicolumn{3}{c|}{\textbf{Single-Machine Out-of-Core}} & \multicolumn{3}{c|}{\textbf{Distributed In-Memory}} & \multicolumn{2}{@{}c@{}|}{\textbf{\;\;Distributed Out-of-Core\;\;}} & \multicolumn{2}{c}{\textbf{GraphMP}} \\ 
\multicolumn{1}{l|}{} & \textbf{GraphChi} & \textbf{X-Stream} & \textbf{GridGraph} & \textbf{Pregel+} & \textbf{PowerGraph} & \textbf{PowerLyra} & \textbf{\;\,GraphD\;\,} & \textbf{\;\,Chaos\;\,} & \textbf{NoCache} & \textbf{Cache} \\ \hline
\textbf{Twitter}  &21.35   &7.66  &14.35   &0.17  &1.11  &0.75  &0.26  &2.66  &0.59  &0.51  \\
\textbf{UK-2007} &64.45   &31.23   &38.07   &1.29  &2.72  &2.49  &1.36  &4.25  &2.49  &2.35  \\
\textbf{UK-2014} &647.58  &697.43  &507.63  &- &- &- &589.64  &371.08  &261.04  &46.50 \\
\textbf{EU-2015} &1627.43   &1478.75   &514.62  &- &- &- &2120.22   &627.23   &320.00  &77.19 \\ \thickhline
\end{tabular}
\end{table*}

\begin{table*}[]
\centering
\caption{Performance comparison between GraphMP with other systems (application: CC; time unit: minutes; time collection: first 10 iterations).}
\label{Tab: perf_all_cc}
\begin{tabular}{@{}c|ccc|ccc|cc|cc@{}}
\thickhline
\multirow{2}{*}{\textbf{Dataset}} & \multicolumn{3}{c|}{\textbf{Single-Machine Out-of-Core}} & \multicolumn{3}{c|}{\textbf{Distributed In-Memory}} & \multicolumn{2}{@{}c@{}|}{\textbf{\;\;Distributed Out-of-Core\;\;}} & \multicolumn{2}{c}{\textbf{GraphMP}} \\ 
\multicolumn{1}{l|}{} & \textbf{GraphChi} & \textbf{X-Stream} & \textbf{GridGraph} & \textbf{Pregel+} & \textbf{PowerGraph} & \textbf{PowerLyra} & \textbf{\;\,GraphD\;\,} & \textbf{\;\,Chaos\;\,} & \textbf{NoCache} & \textbf{Cache} \\ \hline
\textbf{Twitter}  &21.23   &11.78   &16.67   &1.57  &1.39  &1.06  &2.66  &4.85  &0.60  &0.55  \\
\textbf{UK-2007} &61.15   &115.94  &35.82   &7.81  &4.24  &4.07  &16.12   &18.79   &2.82  &2.79  \\
\textbf{UK-2014} &635.97  &1628.00   &533.63  &- &- &- &466.60  &414.76  &219.92  &44.31  \\
\textbf{EU-2015} &1553.70   &2691.37   &867.45  &- &- &- &$2172.95$   &$735.02$   &451.74  &91.25  \\ \thickhline
\end{tabular}
\end{table*}

\subsection{GraphMP vs. GraphChi, X-Stream and GridGraph}

In this set of experiments, we compare the performance of GraphMP with three out-of-core graph processing systems: GraphChi, X-Stream and GridGraph. We do not use VENUS, since it is not open source. We run PageRank, SSSP and CC on Twitter, UK-2007, UK-2014 and EU-2015, and record their processing time of the first 10 iterations and  memory usage.   GraphMP-C denotes the system with compressed edge caching, and GraphMP-NC denotes the system without compressed edge caching.  For fair comparison and simplicity, the first iteration's execution time of each application includes data loading and initialization time.


Table \ref{Tab: perf_all_pagerank}, \ref{Tab: perf_all_sssp} and \ref{Tab: perf_all_cc} show the execution time of each iteration with different systems, datasets and applications. We could observe that GraphMP can considerably improve the graph processing performance, especially when dealing with big graphs. In Graph-C, the performance gain comes from three contributions:  VSW model, selective scheduling, and compressed edge caching.
When running PageRank on EU-2015, GraphMP-NC could outperform GraphChi, X-Steam and GridGraph by 1.21x, 3.58x and 1.46x, respectively. If we enable compressed edge caching, GraphMP-C further improves the processing performance, and outperforms GraphChi, X-Steam and GridGraph by 10.28x, 30.27x and 12.32x to run PageRank on EU-2015, respectively. 
When running SSSP, only a small part of vertices  may update their values in an iteration. With selective scheduling, GraphMP-NC and GraphMP-C could skip loading and processing inactive shards to reduce the disk I/O overhead and processing time.    GraphMP-NC could respectively outperform GraphChi, X-Steam and GridGraph by 5.09x, 4.62x and 1.61x for running SSSP on EU-2015. GraphMP's compressed edge caching mechanism could further reduce disk I/O overhead, and reduce the processing time. Thus, GraphMP-C could outperform GraphChi, X-Steam and GridGraph by 21.08x, 19.16x and 6.67x for running SSSP on EU-2015, respectively. When running CC on EU-2015, GraphMP-NC could outperform GraphChi, X-Steam and GridGraph by 3.42x, 5.96x and 1.92x, respectively. This performance gain is due to the VSW computation model with less disk I/O overhead. If we enable compressed edge caching, GraphMP-C respectively outperforms GraphChi, X-Steam and GridGraph by 17.02x, 29.49x and 9.51x to run CC on EU-2015.

\renewcommand\arraystretch{1}
\begin{table}[]
\centering
\caption{Preprocessing time comparison between GraphChi, GridGraph, X-Stream and GraphMP (time unit: minutes).}
\label{Tab: PreprocessingTime}
\begin{tabular}{c|cccc}
\thickhline
 & \textbf{Graphchi} & \textbf{Gridgraph} & \textbf{X-Stream} & \textbf{GraphMP} \\ \hline
\textbf{Twitter} & 11.08  & 4.83  & 3.38  & 5.67  \\
\textbf{UK-2007} & 45.42  & 23.98  & 14.20  & 20.93  \\
\textbf{UK-2014} & 453.07  & 422.02  & 130.41  & 313.18  \\
\textbf{EU-2015} & 1031.02  & 766.03  & 218.37  & 523.41  \\
\thickhline
\end{tabular}
\end{table}


Table \ref{Tab: PreprocessingTime} shows the data preprocessing time of GraphChi, GridGraph, X-Stream and GraphMP. We use provided data preprocessing programs of GraphChi and GridGraph, and use C++ to implement a new data preprocessing engine for X-Stream, since X-Steam provides a Python script for data preprocessing with poor performance. In this experiment, all input graphs are stored in CSV format. From Table \ref{Tab: PreprocessingTime}, we can find that GraphMP would not introduce much data preprocessing cost. When dealing with EU-2015, X-Stream offers the best performance: it only uses 218.37 minutes to split the input graph into partitions with required format. GraphChi, GridGraph and GraphMP take 1031.02, 766.03 and 523.41 minutes to preprocess EU-2015, respectively.

\setlength{\minipagewidth}{0.235\textwidth}
\setlength{\figurewidthFour}{\minipagewidth}
\begin{figure} 
    \centering
    \begin{minipage}[t]{\minipagewidth}
    \begin{center}
    \includegraphics[width=\figurewidthFour]{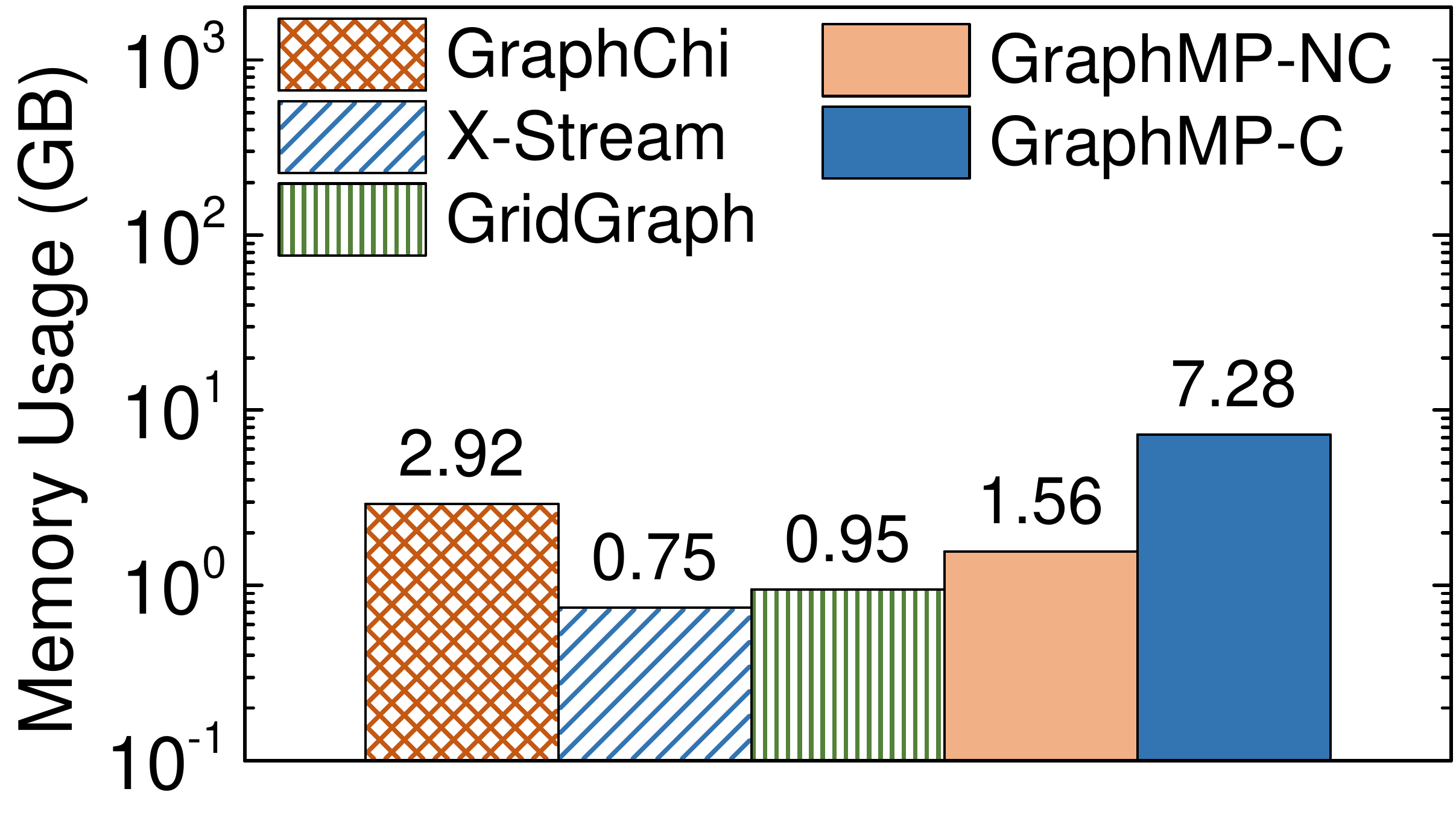}
    \subcaption{(a) Twitter}
    \end{center}
    \end{minipage}
    \centering
    \begin{minipage}[t]{\minipagewidth}
    \begin{center}
    \includegraphics[width=\figurewidthFour]{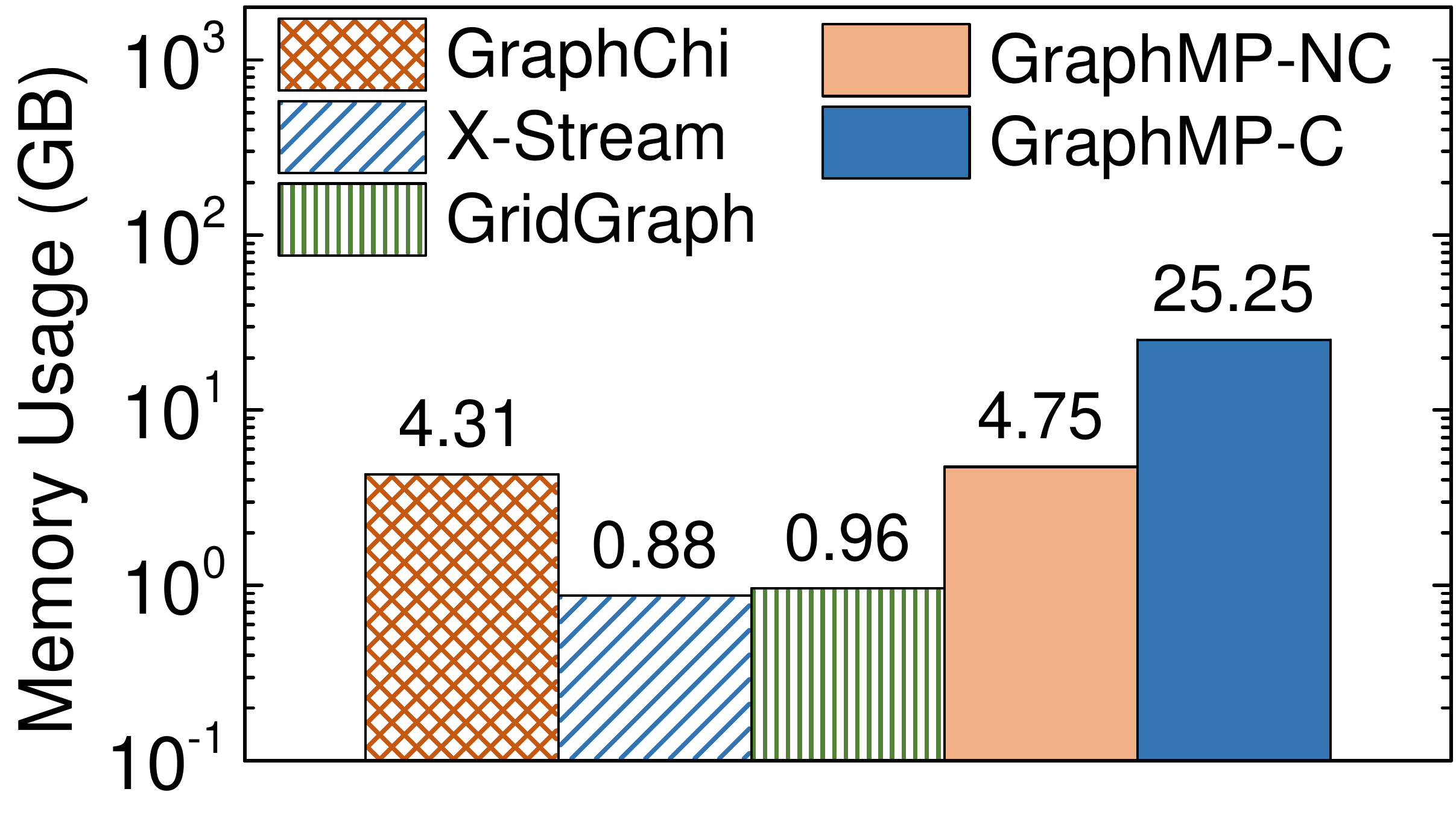}
    \subcaption{(b) UK-2007}
    \end{center}
    \end{minipage}
    \centering
    \begin{minipage}[t]{\minipagewidth}
    \begin{center}
    \includegraphics[width=\figurewidthFour]{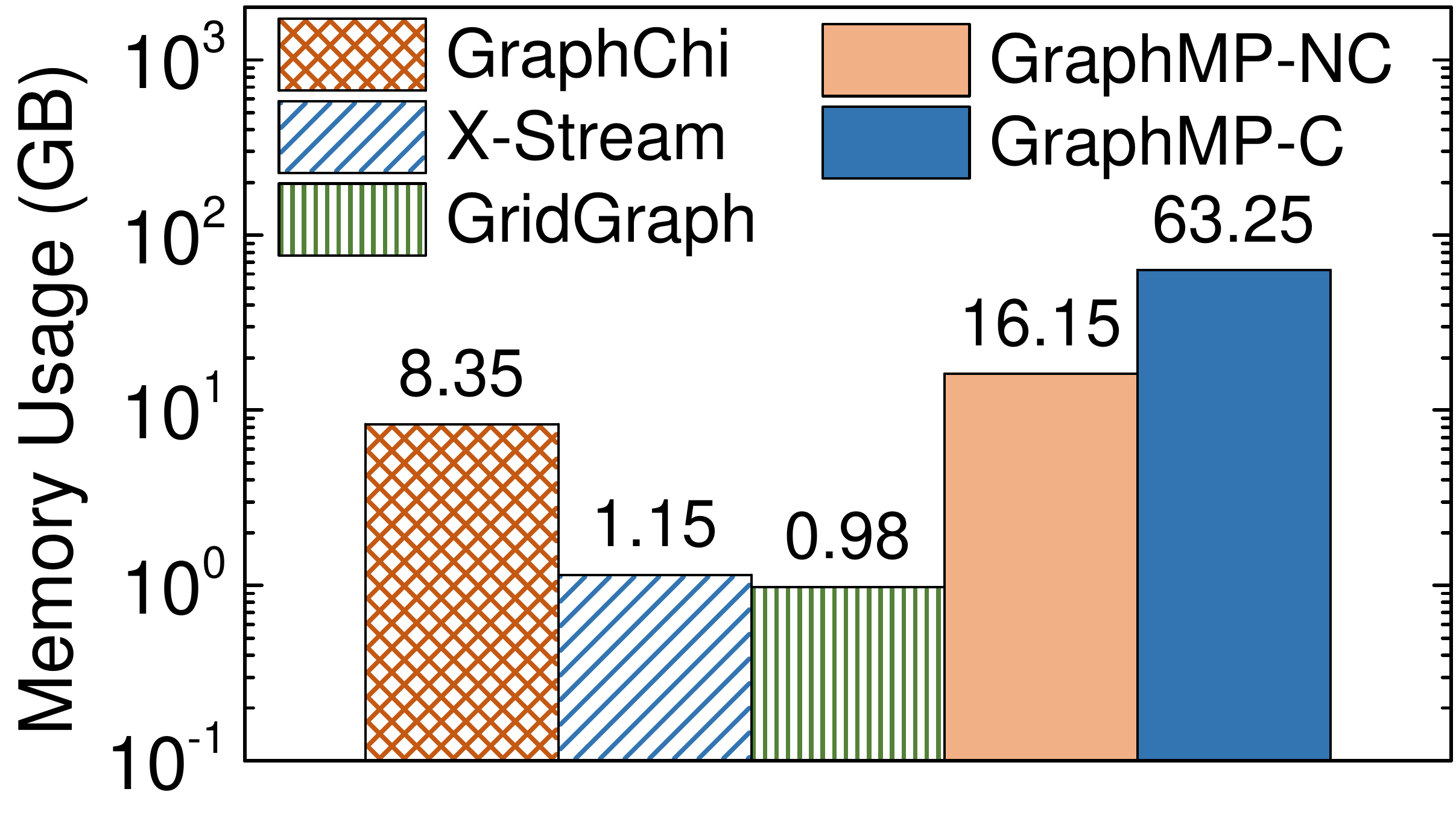}
    \subcaption{(c) UK-2014}
    \end{center}
    \end{minipage}
    \centering
    \begin{minipage}[t]{\minipagewidth}
    \begin{center}
    \includegraphics[width=\figurewidthFour]{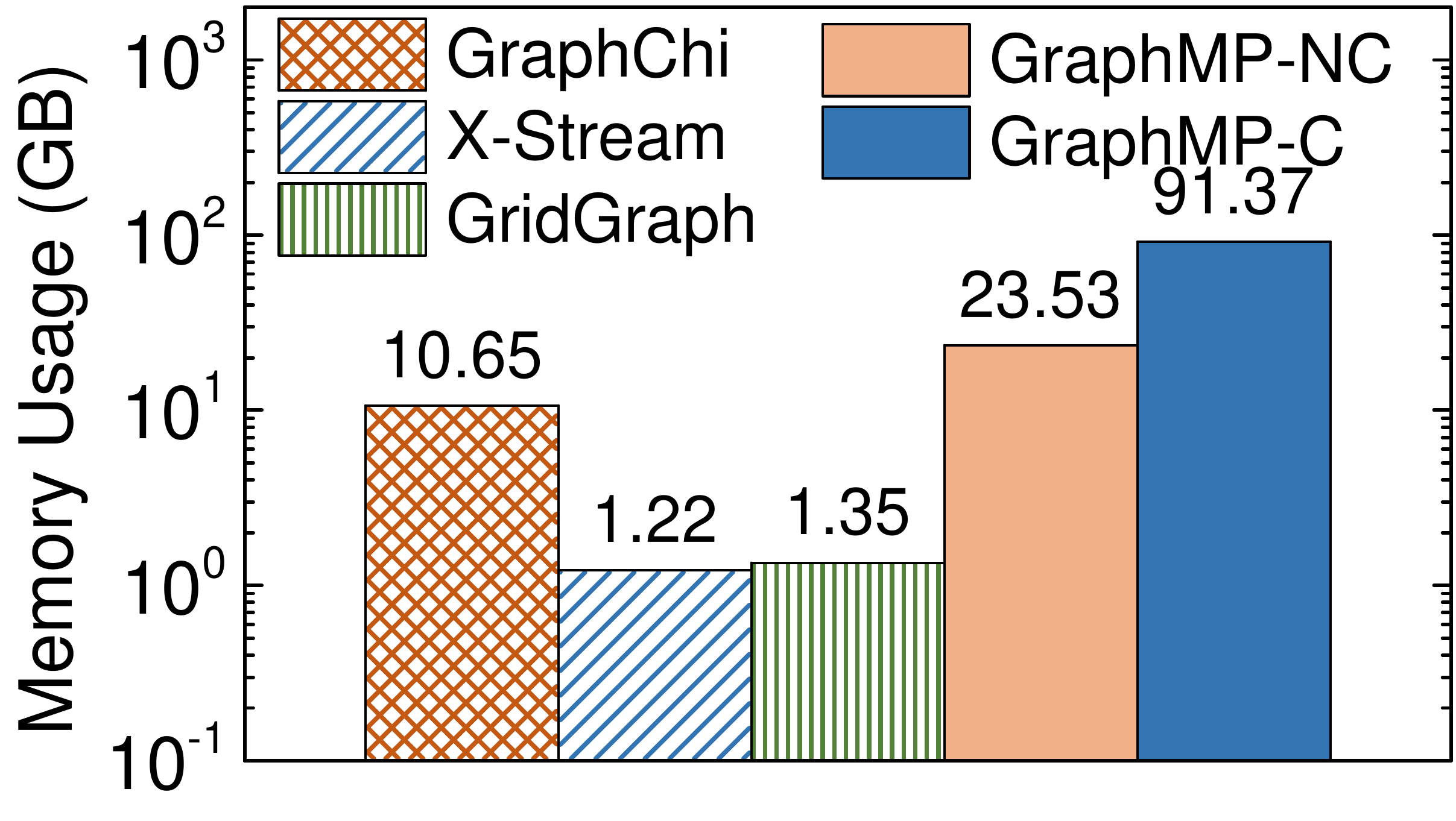}
    \subcaption{(d) EU-2015}
    \end{center}
    \end{minipage}
    \centering
    \caption{Memory usage of 5 graph processing systems to run PageRank on Twitter, UK-2007, UK-2014 and EU-2015. We disable compressed cache mechanism in  GraphMP-NC, and enable it in GraphMP-C.}
\label{Fig: Memory}
\end{figure}


Figure \ref{Fig: Memory} shows the memory usage of each graph processing system to run PageRank. We can see that GraphMP-NC uses more memory than GraphChi, X-Stream and GridGraph, since it keeps all source and destination vertices in memory. For example, when running PageRank on EU-2015, GraphChi, X-Stream and GridGraph only use 10.65GB, 1.22GB and 1.35GB memory, respectively. The corresponding value of GraphMP-NC is 23.53GB. GraphChi, X-Stream and GridGraph are designed for graph processing at scale on a single  PC rather than a commodity server or a cloud instance. Even though our machine has 128GB memory, these systems cannot efficiently utilize them. If we enable compressed edge caching, GraphMP-C uses 91.37GB memory to run PageRank on EU-2015. In this case, GraphMP-C  uses about 68GB as cache. Due to  compression techniques, GraphMP-C could store all 91.8 billion edges in the cache using 68GB memory. Thus, there are no disk accesses for edges during the computation after the first iteration. While  GraphMP-C needs additional time for shard decompression, it can still considerably improve the  processing performance due to the reduced disk I/O overhead.

\subsection{GraphMP vs. Distributed Graph Engines}

We compare the performance of GraphMP with three distributed in-memory graph engines (Pregel+, PowerGraph and PowerLyra) and two distributed out-of-core approaches (GraphD and Chaos). We set up aforementioned distributed graph engines on 9 servers connected by 10Gbps network. Each server has the same hardware and software configuration with the server used to run GraphMP. The cluster totally has $1.15$TB memory and $18$ physical CPUs (108 cores, 216 threads). In the experiments, we enable selective scheduling and compressed edge caching in GraphMP.

Table \ref{Tab: perf_all_pagerank}, \ref{Tab: perf_all_sssp} and \ref{Tab: perf_all_cc} show that GraphMP can be as highly competitive as distributed graph processing systems. When running PageRank on UK-2007, GraphMP outperforms Pregel+ and PowerGraph by $1.89$x and $1.11$x, respectively. In the same case, PowerLyra  outperforms GraphMP by $1.13$x. Compared to GraphD and Chaos, GraphMP could speed up PageRank on UK-2007 by a factor of $4.61$ and $4.86$. When running SSSP on UK-2007, Pregel+ and GraphD offer better performance than GraphMP, since they could avoid processing inactive vertices at the level of vertex. As a comparison, GraphMP could only perform selective scheduling at the level of edge shards. Compared to Chaos, GraphMP could speed up SSSP on UK-2007 by a factor of $1.81$. When running CC on UK-2007, GraphMP could outperform Pregel+, PowerGraph, PowerLyra, GraphD and Chaos by $2.79$x, $1.52$x, $1.46$x, $5.78$x and $6.74$x, respectively. Note that these distributed graph engines have 9x more resources than GraphMP. 

Due to memory limitation (even though the cluster has more than 1TB memory), Pregel+, PowerGraph and PowerLyra crash when processing UK-2014 and EU-2015. GraphD and Chaos can cope with UK-2014 and EU-2015 from disks. As a comparison, GraphMP can efficiently handle UK-2014 and EU-2015 using just a single machine. With compressed edge caching, GraphMP can manage all edges in a single machine's memory. Therefore, GraphMP avoids costly disk I/O operations, and offers higher performance than GraphD and Chaos. Compared to GraphD, GraphMP could speed up the processing performance by a factor of $23.99$, $27.46$ and $23.81$ to run PageRank, SSSP and CC on EU-2015. Compared to Chaos, the corresponding speedup ratios are $7.95$, $8.13$ and $8.06$, respectively.

\section{Conclusion}
 
In this paper, we tackle the challenge of big graph analytics on a single machine. Existing out-of-core approaches have poor  performance due to the high disk I/O overhead. To address this problem, we propose a new out-of-core graph processing system named GraphMP. GraphMP partitions the input graph into small shards, each of which could be fully loaded into memory and contains a similar number of edges. Edges with the same destination vertex appear in the same shard. We use three techniques to improve the graph processing performance by reducing the disk I/O overhead. First, we design a vertex-centric sliding window  computation model to avoid reading and writing vertices on disk. Second, we propose selective scheduling to skip loading and processing unnecessary shards on disk. Third, we use compressed edge caching to fully utilize the available memory resources to reduce the amount of disk accesses for edges. With these three techniques, GraphMP could efficiently support big graph analytics on a single commodity machine. Extensive evaluations show that GraphMP could outperform GraphChi, X-Stream and GridGraph by up to 30, and can be as highly competitive as distributed graph engines like Pregel+, PowerGraph and Chaos. 

GraphMP is designed for graph processing on normal machines or cloud instances without special hardwares. If large flash memory or non-volatile memory is deployed, one may use  systems like FlashGraph or Mosaic. When having big memory (for example, in supercomputers), one may use in-memory systems like GraphMat. If high-speed network and big memory machines are available, one may use distributed graph engines like Pregel+ or PowerGraph.

\bibliographystyle{IEEEtran}
\bibliography{main}

\end{document}